\documentclass[fontsize=11pt,parskip=yes,onecolumn]{scrartcl}
\usepackage{graphicx}
\graphicspath{ {./img/} }
\usepackage[
figurename=Figure,
font={sf,normalsize},
format=plain,
labelfont=bf,
]{caption}
\usepackage[
natbib=true,
backend=bibtex,
maxcitenames=1,
minbibnames=3,
maxbibnames=40
]{biblatex}
\usepackage{float}
\usepackage{makecell} %multiple line in a table cell
\usepackage{pbox} %multiple line in a table cell
\usepackage{subcaption}
\usepackage{xcolor}
\usepackage{soul} %highlight text
\newcommand{\abs}[1]{\left\vert #1 \right\vert}

\usepackage{indentfirst} %indent after a section
\usepackage{amsmath}
\usepackage{amssymb}
\usepackage{geometry}
\usepackage[utf8]{inputenc}
\usepackage[section]{placeins} %placing section (Ref) at the bottom
\usepackage{hyperref}

\usepackage{amsmath}
\usepackage{tikz}
\usetikzlibrary{matrix}
\usetikzlibrary{decorations.pathreplacing}
\usepackage{dsfont}
\usepackage{tabularx,ragged2e,booktabs}
\usepackage{authblk}
\usepackage{tcolorbox} % create box
\usepackage{algorithm}
\usepackage{algorithmic}

\usepackage[shadow,loadshadowlibrary]{todonotes}
\definecolor{purple1}{rgb}{0.949, 0.941, 0.969}
\definecolor{purple5}{rgb}{0.329, 0.153, 0.561}
\definecolor{green1}{HTML}{edf8fb}
\definecolor{green5}{HTML}{006d2c}
\definecolor{pink1}{HTML}{f1eef6}
\definecolor{pink4}{HTML}{dd1c77}
\definecolor{blue1}{HTML}{D6EAF8}
\definecolor{blue2}{HTML}{3357FF}
\usepackage{blindtext}

\usepackage{orcidlink}
\usepackage{setspace}
\title{Spatial dynamic modelling to understand how dendritic cell clustering affects T cell activation}
\author[1,3]{Domenic P.J. Germano \orcidlink{0000-0001-5893-4840}}
\author[2]{Federico Frascoli \orcidlink{0000-0003-4159-6229}}
\author[3]{Robyn P. Araujo \orcidlink{0000-0002-3360-2214}}
\author[4]{Peter P. Lee \orcidlink{0000-0002-2660-4377}}
\author[1]{Peter S. Kim \orcidlink{0000-0002-1492-4744}}

\affil[1]{The School of Mathematics and Statistics, The University of Sydney, Camperdown, NSW, Australia}
\affil[2]{Department of Mathematics, School of Science, Computing and Engineering Technologies, Swinburne University of Technology, Hawthorn, Vic, Australia}
\affil[3]{The School of Mathematics and Statistics, The University of Melbourne, Parkville, Vic, Australia}
\affil[4]{Department of Immuno-Oncology, City of Hope and Beckman Research Institute, Duarte, CA, USA}

\date{}    %% if you don't need date to appear
\begin{document}
% \linenumbers % Start numbering here
\maketitle

\begin{abstract}
% The adaptive immune system is responsible for maintaining the body's healthy cell population.
% This maintenance involves the priming and activation of immune cells, such as T cells, towards various different cell types. T cells require stimulation towards specific antigens to activate, a process which occurs within the lymph nodes. T cell stimulation is achieved via interaction with antigen presenting Dendritic cells. Ensuring that T cell priming occurs rapidly and robustly is a key feature of healthy cohorts of patients.
The coordination of the immune system and its components is essential for the body to maintain a healthy status.
Recent clinical studies show that breast cancer patients with high Dendritic cell clustering in tumour draining lymph nodes have improved survival outcomes, compared to those with a lower degree of clustering. These results suggest that a specific form of Dendritic cell clustering  promotes T cell activation.
However, the mechanistic effects of this spatial organisation is unclear.

We develop a spatially dynamic model of T cells interacting with Dendritic cells within the lymph node. We present a novel probabilistic agent-based model (ABM) of T cells, and use it to derive the deterministic, phenotypically structured partial differential equation (PS-PDE) of T cell activation and motion. Using the PS-PDE, we derive analytic approximations of the expected T cell stimulation distribution, based on the topology and level of clustering of a given Dendritic cell population. Our analytic approximation enables us to identify T cell characteristics that benefit most from Dendritic cell clustering, to result in an enhanced stimulation distribution. We also perform a sensitivity analysis with our models to identify T cell characteristics that result in desirable T cell activation characteristics, such as rapid T cell activation, and robust heterogeneous T cell activation.

Our key findings show that T cells with an intermediate level of stimulation uptake benefit most from higher levels of Dendritic cell clustering, activating with a comparable or greater abundance, and greater heterogeneity, when compared to T cells of a similar characteristic but with a lower level of Dendritic cell clustering.
\end{abstract}


% \section*{Author Summary}
% The adaptive immune system maintains immunity in a specialised, targeted fashion. Dendritic cells (DCs) play a central role in this process, activating T cells that identify and eliminate abnormal or cancerous cells. Recent clinical studies in breast cancer patients have shown that those whose tumour draining lymph nodes contain highly clustered DCs tend to have better survival outcomes compared to patients with lower DC clustering, suggesting that the spatial organisation of DCs directly influences the effectiveness of the T cell response.
% In this study, we use mathematical modelling to investigate how DC clustering affects T cell activation within the lymph node. Using our model, we show that T cells with an intermediate capacity for stimulation benefit most from DC clustering, becoming activated in greater numbers and with greater heterogeneity. This work has broad clinical implications towards T cell properties that may benefit most from DC clustering.

\section{Introduction}
\begin{figure}[h!]
    \centering
    \includegraphics[width=0.9\linewidth]{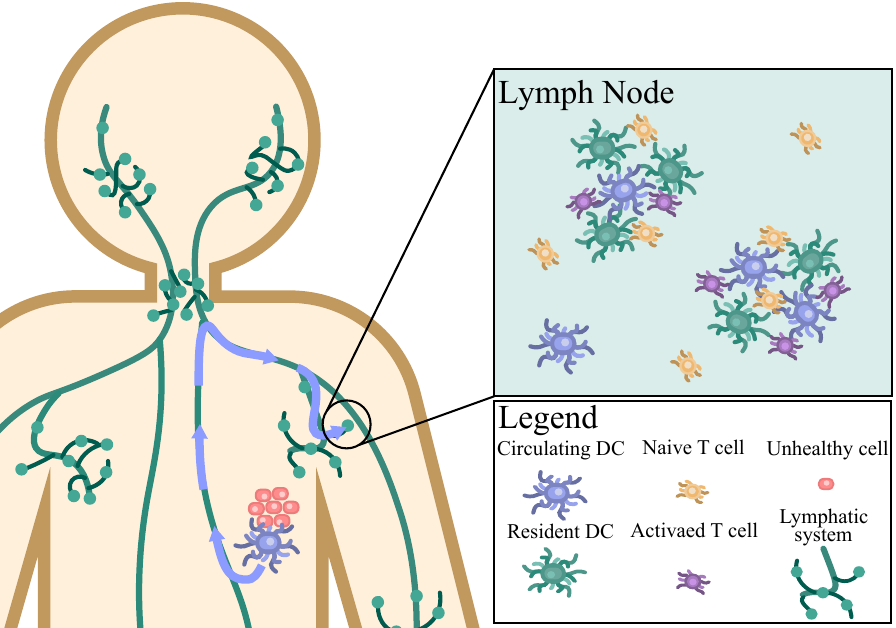}
    \caption{An illustration of circulating dendritic cells (DCs) collecting antigen specific to the unhealthy cells, and being transported to the lymph node, via the lymphatic system. Within the lymph node, the circulating DCs interact with the resident DCs, and also naïve T cells. Naïve T cells then become activated towards a particular antigen, becoming either CD4\textsuperscript{+} helper T cells or CD8\textsuperscript{+} cytotoxic T cells to then clear the unhealthy cells (not shown).}
    \label{fig:ProblemSchematic}
\end{figure}

The immune system comprises of two key branches: the innate immune system, and the adaptive immune system. Both the innate and adaptive systems are tightly coupled to help keep the body healthy \cite{iwasaki2015control}. While the innate immune system acts as a rapid, non-specific first line of defence, via recognising and responding to a broad range of pathogens,  the adaptive immune response is slower but highly specific, targeting particular antigens and retaining immunological memory for more effective future responses \cite{ iwasaki2010regulation, kennedy2010brief, tomar2014brief}. Immune maintenance entails the identification of abnormal, unhealthy cells. To identify these abnormal cells, dendritic cells (DCs) are continually transported throughout the body to collate and gather information about the body's cellular composition \cite{bonasio2006generation, philip2022cd8}. Upon encountering abnormal cells, these DCs will gather antigens specific to these abnormal cells, as a means to identify them \cite{worbs2017dendritic}. The migratory DCs are then transported to a local draining lymph node, where they interact with the resident DC population and exchange the accumulated antigen \cite{bonasio2006generation, platt2013dendritic}. These resident DCs then present these antigen (on their surface) to naïve, undifferentiated T cells \cite{randolph2005dendritic}, which are believed to facilitate  T cell recruitment via chemokines secretion \cite{castellino2006chemokines, beuneu2006cutting}. Upon activation towards a particular antigen, T cells proliferate and mature into effector and memory cells, with CD4\textsuperscript{+}  helper T cells and CD8\textsuperscript{+} cytotoxic T cells  \cite{pishesha2022guide, buchholz2016t}. CD4\textsuperscript{+} helper T cells are mainly responsible for the downstream activation of subsequent immune cells, such as B cells, or cytotoxic T cells \cite{pishesha2022guide}. CD8\textsuperscript{+} cytotoxic T cells are directly responsible for the killing of abnormal unhealthy cells \cite{pishesha2022guide}. Therefore, ensuring that both helper and cytotoxic T cell activation occurs in an efficient and effective manner is central for the immune system to ensuring the body is maintained in a healthy state. Figure 1 shows an illustration of the key dynamics.
Figure \ref{fig:ProblemSchematic} shows an illustration of the key dynamics.

% In a recent retrospective tissue-based study, tumour draining lymph nodes (TDLNs) from a cohort involving 59 breast cancer patients were analysed and compared to healthy intra-mammary lymph nodes (HLNs) \cite{chang2013spatial}.
% The authors observed that degree of DC clustering in TDLNs was reduced, when compared to HLNs. Within TDLNs, DC clusters were found to contain fewer DCs and also fewer cells displaying DC maturity marker CD83.
% It was also observed that more numerous T cells were found closer to clustered DCs compared to unclustered DCs, concluding that DC clustering is implicated in promoting DC-T cell interaction.
% Most importantly, the authors found a correlation between DC clustering in TDLNs and the duration of disease-free survival in breast cancer patients \cite{chang2013spatial}, which is highly promising.
In a landmark tissue-based study, tumour draining lymph nodes (TDLNs) from a cohort involving 59 breast cancer patients were analysed and compared to healthy intra-mammary lymph nodes (HLNs) \cite{chang2013spatial}. The authors observed that degree of DC clustering within TDLNs was reduced, when compared to HLNs. Within TDLNs, DC clusters were found to contain substantially fewer DCs and also fewer cells displaying the DC maturity marker CD83. It was also observed in HLNs that more numerous T cells were found closer to clustered DCs compared to unclustered DCs, concluding that DC clustering is implicated in promoting DC-T cell interaction. Most importantly, the authors found a strong correlation between DC clustering in TDLNs and the duration of disease-free survival in breast cancer patients \cite{chang2013spatial}.

% Mathematical models involving the interactions between immune cells and unhealthy cells have attracted lots of interest in mathematical biology \cite{almeida2022hybrid, castro2016mathematics, eftimie2016mathematical}. 
% Developing models to provide insights into how T cells become activated towards a particular phenotype has also gained much attention \cite{riggs2008comparison, kumbhari2020mature, gong2014harnessing}. 
% There have also been many descriptions of the spatial interactions between DCs and T cells, in the context of T cell activation \cite{vroomans2012chemotactic, azarov2019role, johnson2021lymph, moreau2016virtual, textor2014random,banigan2015heterogeneous}.
% However, to date, we have not found a mathematical model that describes the degree of spatial organisation of immune cells and implications to their activation.
% Therefore, in this work, we use mathematical models to allow us to better understand how the spatial organisation of the DCs within the lymph nodes may result in improved T cell activation.
% To achieve this, we develop a discrete, agent-based model (ABM) of T cell movement, and then derive a continuum, phenotype-structured partial differential equation (PS-PDE) model based on the aforementioned discrete model. This work builds upon the emerging work of structured PDEs in the context of biological cells \cite{lorenzi2025phenotype,morselli2024phenotype}. In both models, the T cell population interacts with a discrete dendritic cell population.
Mathematical models involving the interactions between immune cells and unhealthy cells have attracted lots of interest in mathematical biology \cite{almeida2022hybrid, castro2016mathematics, eftimie2016mathematical}. Developing models to provide insights into how T cells become activated towards a particular phenotype has also gained much attention \cite{riggs2008comparison, kumbhari2020mature, gong2014harnessing}. There have also been many descriptions of the spatial interactions between DCs and T cells, in the context of T cell activation \cite{vroomans2012chemotactic, azarov2019role, johnson2021lymph, moreau2016virtual, textor2014random,banigan2015heterogeneous}. However, to date, we have not found a mathematical model that describes the degree of spatial organisation of immune cells and implications to their activation. Therefore, in this work, we use mathematical models to allow us to better understand how the spatial organisation of the DCs within the lymph nodes may result in improved T cell activation. To achieve this, we develop a discrete, agent-based model (ABM) of T cell movement, and then derive a continuum, phenotype-structured partial differential equation (PS-PDE) model based on the aforementioned discrete model. This work builds upon the emerging work of structured PDEs in the context of biological cells \cite{lorenzi2025phenotype,morselli2024phenotype}. In both models, the T cell population interacts with a discrete dendritic cell population.

% The remainder of the paper is structured as follows. We provide the details of the mathematical models, describing the DC population and how they attract T cells towards them, and also how the DCs are clustered into groups of various sizes. We describe the T cell motion and activation via interactions with DCs for the ABM and propose the PS-PDE model (for a formal derivation, see \ref{si:cont_derivation}). We then provide a model comparison between the ABM and the PS-PDE for a simple 1D domain and a single DC in 2D. The key results are presented in the final results section, utilising both modelling approaches where appropriate, focusing on how T cell activation varies with DC cluster size. Lastly the final remarks and possible implications for our work is discussed in the final section. 
% , applied to understand the implications for T cell activation in the context of varying DC cluster size.
The remainder of the paper is structured as follows. We provide the details of the mathematical models, describing the DC population and how they attract T cells towards them, and also how the DCs are clustered into groups of various sizes. We describe the T cell motion and activation via interactions with DCs for the ABM and propose the PS-PDE model (for a formal derivation, see \ref{si:cont_derivation}). We then provide a model comparison between the ABM and the PS-PDE for a simple 1D domain and a single DC in 2D. The key results are presented in the final results section, utilising both modelling approaches where appropriate, focusing on how T cell activation varies with DC cluster size. Lastly the final remarks and possible implications for our work is discussed in the final section.

\section{Mathematical Models of T cell dynamics and DC interaction}
In this section, we describe the mathematical modelling techniques used to describe the system. 
We represent the static dendritic cell population by discrete points in space. To model T cell movement, we use two different approaches: a discrete, agent-based model (ABM) and a continuum, phenotype structured partial differential equation (PS-PDE) model. In the following sections, we first describe the model for the DC population and each of the T cell models. We then present a steady-state, analytical approximation to the long term behaviour of the T cell stimulation distribution, using the continuum model.

\subsection{Dendritic cell spatial organisation}
\begin{figure}[h!]
    \centering
    \begin{subfigure}[b]{0.3\textwidth}
    \centering
    \includegraphics[width = 0.9\linewidth]{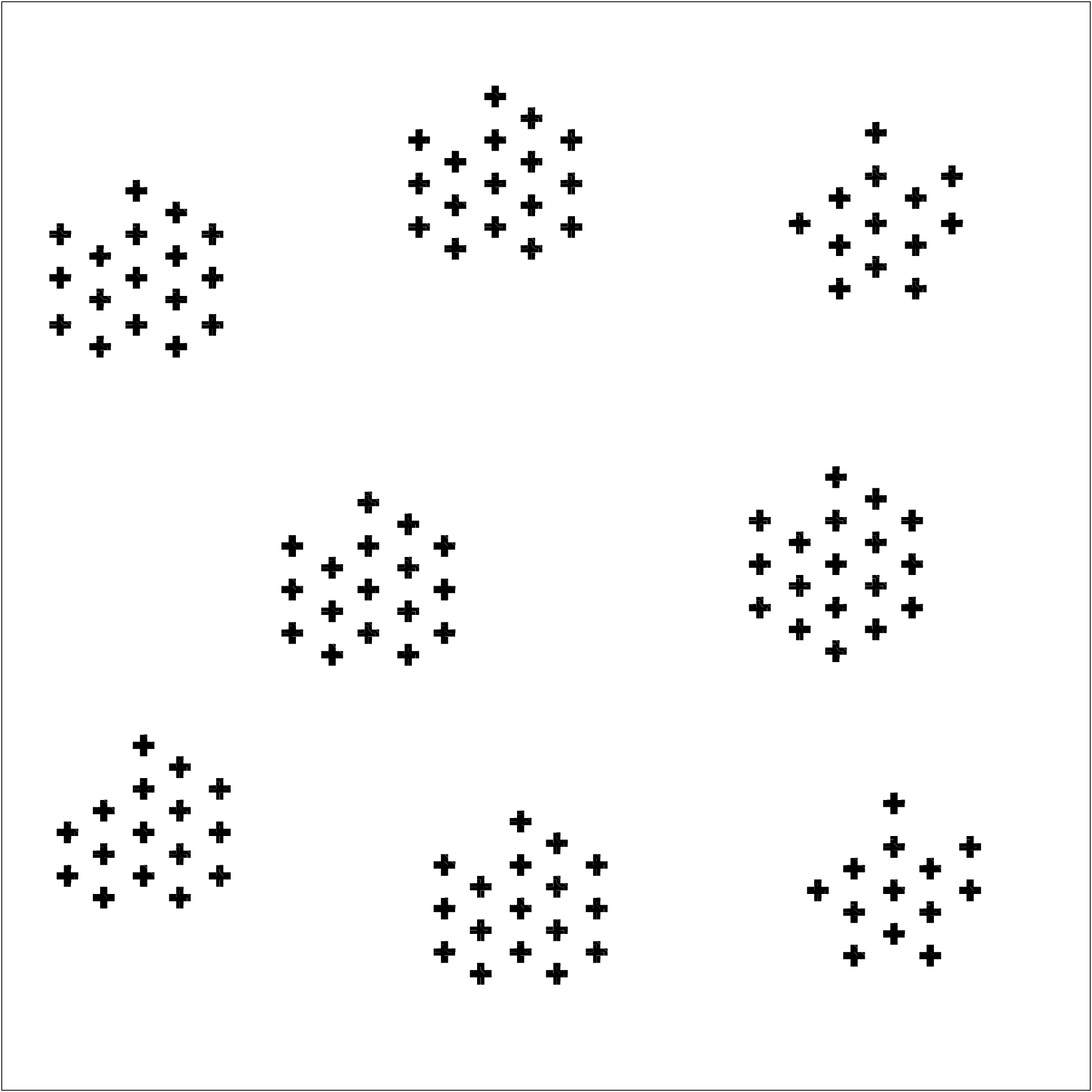}
    \caption{\label{fig:cluster_8}}
    \end{subfigure}
    \begin{subfigure}[b]{0.3\textwidth}
    \centering
    \includegraphics[width = 0.9\linewidth]{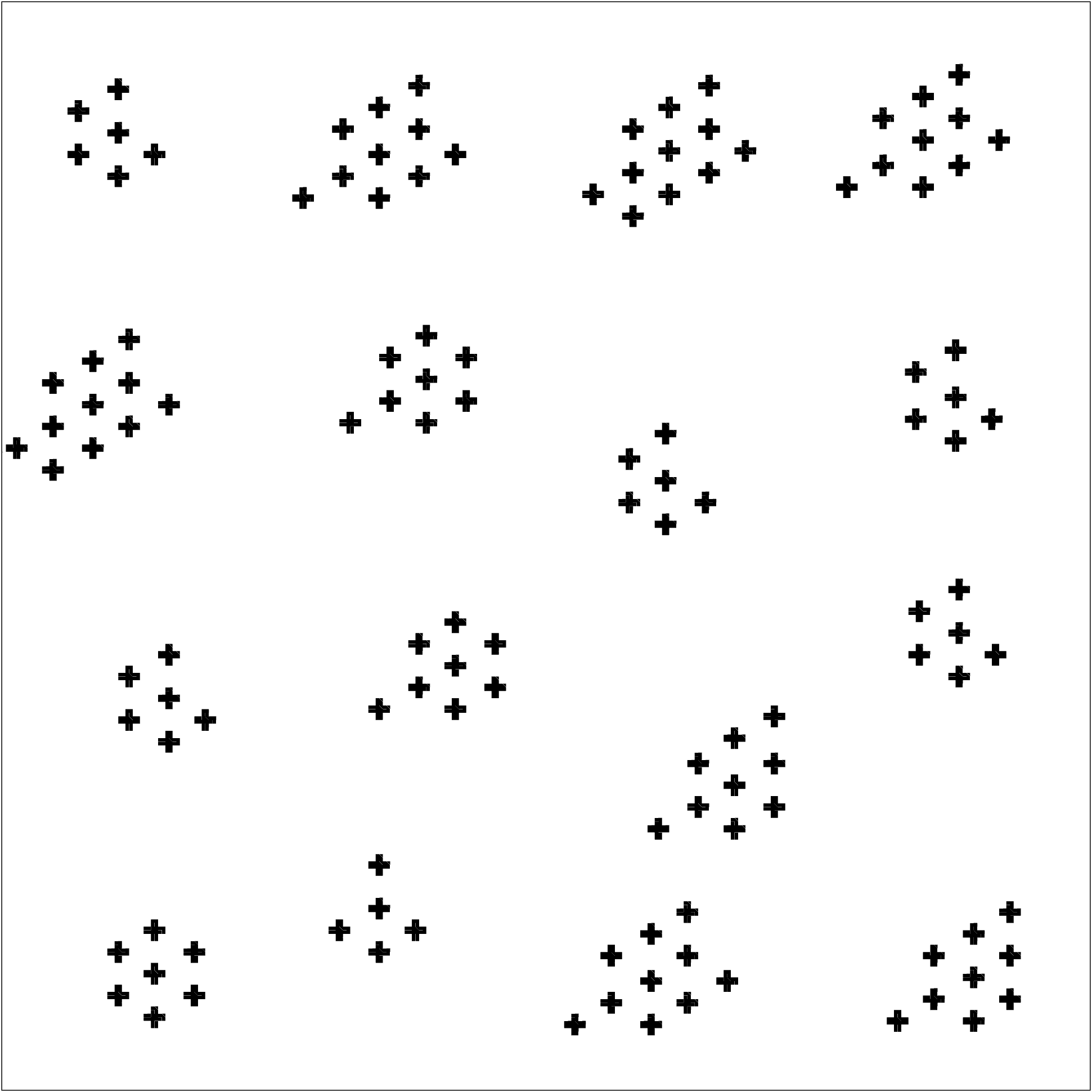}\caption{\label{fig:cluster_4}}
    \end{subfigure}
    \begin{subfigure}[b]{0.3\textwidth}
    \centering
    \includegraphics[width = 0.9\linewidth]{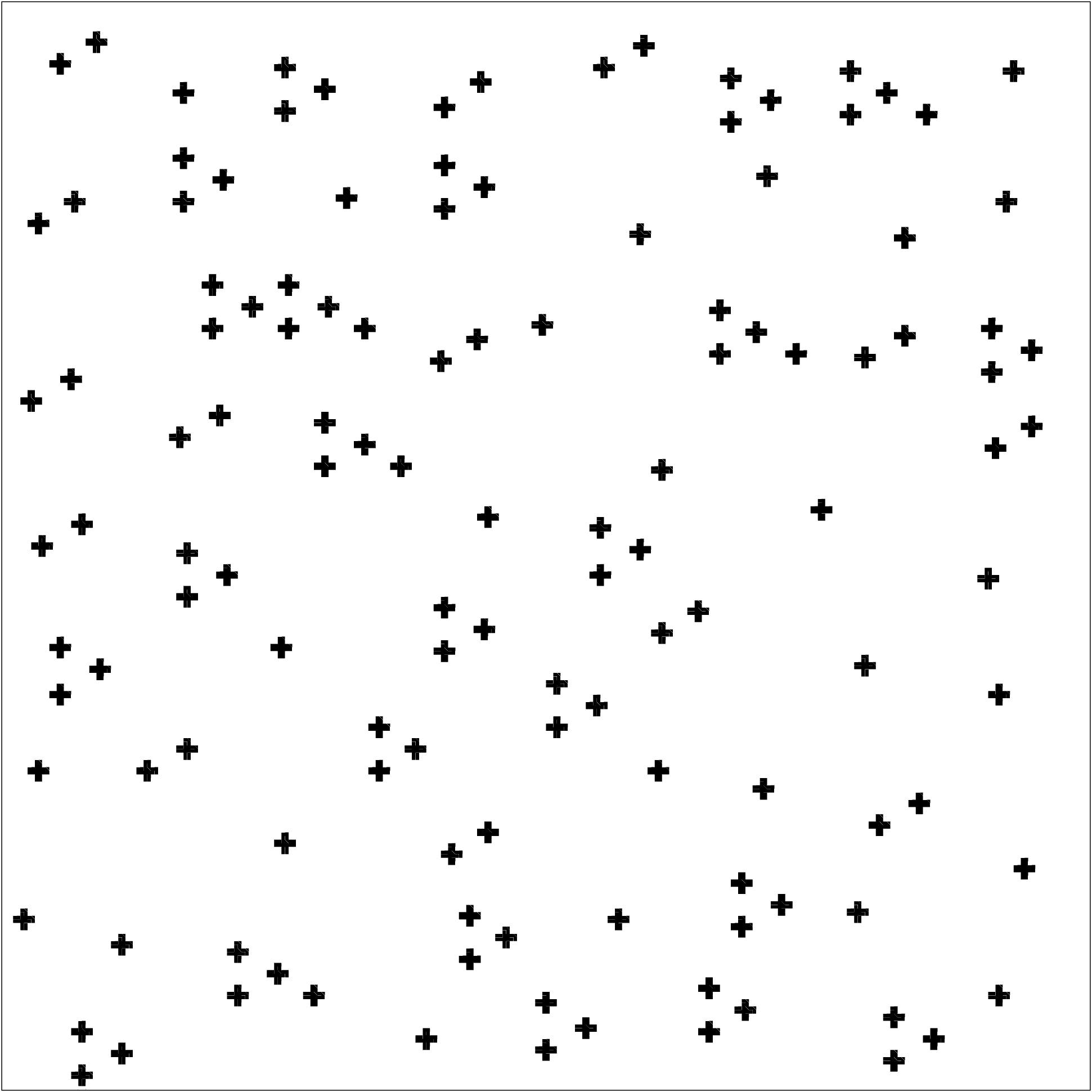}
    \caption{\label{fig:cluster_2}}
    \end{subfigure}

    \captionsetup{subrefformat=parens}
    \caption{Three examples of Dendritic cell (DC) clustering with fixed DC density (128 DCs), in a tissue of size $150\times150$ T cell diameters. \ref{fig:cluster_8} is an example with mean clusters sizes of $K_m = 16$, resulting in DC 8 clusters, \ref{fig:cluster_4} is an example  with mean clusters sizes of $K_m = 8$, resulting in DC 16 clusters, and \ref{fig:cluster_2} is an example  with mean clusters sizes of $K_m = 2$, resulting in 64 DC clusters.}
    \label{fig:DC_Cluster}
\end{figure}

We first outline how we represent the characteristics affecting dynamics of the dendritic cell (DC) population within the lymph node.
There are three main components to describe the DCs: (i) their shape, (ii) position in space, $\mathbf{r}_k = (x_k,y_k)$, and (iii) a chemokine secreted to attract T cells. The chemokine is produced at rate $s$ by each dendritic cell, which is assumed to originate from $\mathbf{r}_k$ for each cell $k$, decays at rate $K$, and diffuses with diffusivity $D_C$. The chemokine concentration, $C_k$ due to dendritic cell, $k$ is given by the following expression:
\begin{align}
    \frac{\partial C_k}{\partial t} = D_C \nabla^2 C_k + s \delta(\mathbf{r}-\mathbf{r}_k)- K C_k.
\end{align}
We assume that chemokines are produced and diffuse at a much quicker rate, compared to that at which T cells consume and respond to them. Therefore, we say that the chemokine concentration is at a steady-state and with negligible consumption by T cells, resulting in the solution:
\begin{align}
    C_k(\mathbf{r}) = \frac{1}{2 \sqrt{D_C K}} \exp{\left( - \abs{\mathbf{r}-\mathbf{r}_k} \sqrt{ \frac{K}{D_C} }\right)}.
\end{align}
If we have the chemokine diffusing throughout the lymph node without obstruction, the total level of chemokine within the lymph node $C$ is given as the superposition of $C_k$:
\begin{align}\label{eq:chemokine_field}
    C(\mathbf{r}) = \kappa \sum_{\forall k} C_k(\mathbf{r}),
\end{align}
with $\kappa$ chosen to ensure the maximum chemokine concentration, $C$, is unitary. 
DCs are modelled as plus-shaped symmetric stencils, spanning 3 units wide. We define $\mathds{1}_{\text{DC}}$ as the indicator function for the region containing DCs, i.e.:
\begin{align}
    \mathds{1}_{\text{DC}}(x,y) = 
    \begin{cases}
        1, \quad \text{for } (x,y) \in \Omega_{\text{DC}},\\
        0, \quad \text{for } (x,y) \notin \Omega_{\text{DC}},
    \end{cases}
\end{align}
where the domain $\Omega_{\text{DC}}$ is the domain containing DCs.
These cells are distributed throughout the domain in $K_{DC}$ local clusters with a mean cluster size of $K_m$, that are organised in a hexagonal-like configuration. In each cluster, DCs are placed from the centre location outwards, with space between the cells to allow for T cells to move between adjacent DCs. To ensure variation in cluster configurations, a ring may contain empty unoccupied positions, with outer rings being populated instead. Three examples of the dendritic cell clustering are given in Figure \ref{fig:DC_Cluster} for mean clusters sizes of $K_m = 16$ (\ref{fig:cluster_8}), $K_m = 8$ (\ref{fig:cluster_4}), and $K_m = 2$ (\ref{fig:cluster_2}). These spatial arrangements and their variations capture prototypical structures present in real-life dendritic cell clustering \cite{chang2013spatial, bousso2003dynamics, huang2026dendritic}.

\subsection{T cell motion and activation}
We describe T cell movement via diffusion and chemotaxis, which, for our discrete model, each define a set of rules, leading to a discrete-time, biased random walk \cite{krummel2016t, banigan2015heterogeneous, celli2012many}. We also relax space-filling restrictions, so that more than one cell may occupy a lattice site; however, we only consider a fixed T cell population where cells do not proliferate. Finally, T cells gain stimulation based on their proximity to a dendritic cell.
% Accounting for the long-range of the antigen-presenting dendrites on the dendritic cells, when a T cell is at least 1 unit away from any part of the plus-shaped dendritic cell, the T cell may gain antigen. If a T cell moves further than 1 unit from the dendritic cell, then the accumulated antigen may degrade, resulting in antigen loss.
Accounting for the long-range of the antigen-presenting dendrites on the DCs, when a T cell is at least 1 unit away from any part of the plus-shaped dendritic cell, the T cell becomes stimulated towards activating. If a T cell moves further than 1 unit from the dendritic cell, then the accumulated stimulation may degrade, resulting in stimulation loss, mimicking actual processes in T-cell activation \cite{friedman2010real}.
Once a T cell has accumulated a sufficient amount of stimulation, defined as $A_{\text{max}}$, the T cell is activated and can no longer accumulate or lose stimulation.

\subsection{Discrete model}
The discrete, agent-based model of the T cells is described as a discrete-time, biased random walker in a domain with reflective boundary conditions.
We prescribe a time-step, $\tau \in \mathbb{R_{+}}$, and spatial-discretisation, $\delta \in \mathbb{R}_{+}$, and define $x_i := i \delta$, $y_j: = j\delta$, and $t_k := k \tau $.
T cells cannot pass through DCs, and so we impose reflective boundaries on the surface of the DCs. 
T cells first move via an undirected random walk, with probability $0 \leq \theta \leq 1$. 
Therefore, at time-step $k$, if a T cell is located at position $(x_i,y_j)$, then the probability the T cell moves left ($R_{\text{L}}$), right ($R_{\text{R}}$), up ($R_{\text{U}}$), down ($R_{\text{D}}$), or stays still ($R_{S}$) is given, respectively, as:
\begin{equation} \label{eq:rand_prob}
\begin{split}
    R^{k}_{\text{L}}(x_i,y_j) &=  \frac{\theta}{4}\left[1-\mathds{1}_{\text{DC}}(x_{i-1},y_j) \right],\\
    R^{k}_{\text{R}}(x_i,y_j) &=  \frac{\theta}{4}\left[1-\mathds{1}_{\text{DC}}(x_{i+1},y_j) \right],\\
    R^{k}_{\text{U}}(x_i,y_j) &=  \frac{\theta}{4}\left[1-\mathds{1}_{\text{DC}}(x_i,y_{j+1}) \right],\\
    R^{k}_{\text{D}}(x_i,y_j) &=  \frac{\theta}{4}\left[1-\mathds{1}_{\text{DC}}(x_i,y_{j-1}) \right],\\
    R^{k}_{\text{S}}(x_i,y_j) &=  \left(1 - \theta \right)\mathds{1}_{\text{DC}}(x_i,y_j).
\end{split}
\end{equation}

\begin{figure}[h!]
    \centering
    \includegraphics[width=0.25\linewidth]{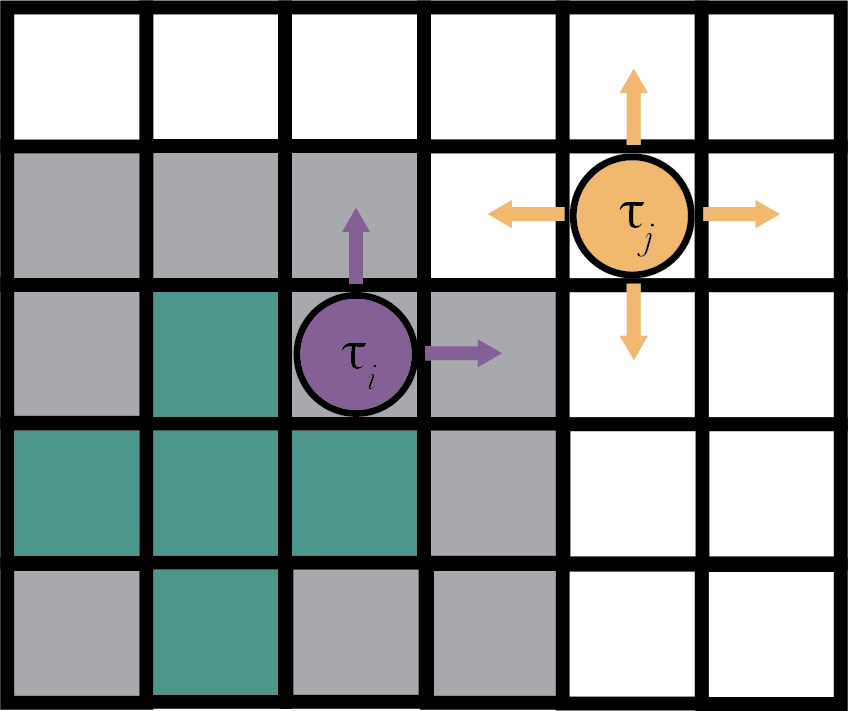}
    \caption{Discrete model illustration, depicting allowable T cell movements. The purple T cell ($i$) is only permitted to move up or right, due to the boundary imposed by the green plus-shaped dendritic cell. The yellow T cell ($j$) is permitted to move in each of the four directions. Furthermore, T cells close to DCs (those in the grey region, such as the purple T cell) can accumulate stimulation from the DCs. T cells away from DCs (those in the white region, such as the yellow T cell) can lose stimulation.}
    \label{fig:discrete_model_illustration}
\end{figure}

Following the undirected random walk, T cells undergo a chemotactic step, with chemotaxis sensitivity $\phi \geq 0$. At time-step $k$, a T cell located at position $(x_i,y_j)$, will move via chemotaxis in a given direction proportional to the change in chemokine gradient in that direction. That is, the probability of moving left ($T_{\text{L}}$), right ($T_{\text{R}}$), up ($T_{\text{U}}$), down ($T_{\text{D}}$), or staying still ($T_{S}$) is given, respectively, as:
\begin{equation} \label{eq:taxis_prob}
\begin{split}
    T^{k}_{\text{L}}(x_i,y_j) &=  \phi \left[\frac{C(x_{i-1},y_j)-C(x_i,y_j)}{4}\right]\left[1-\mathds{1}_{\text{DC}}(x_{i-1},y_j) \right],\\
    T^{k}_{\text{R}}(x_i,y_j) &= \phi \left[\frac{C(x_{i+1},y_j)-C(x_i,y_j)}{4}\right]\left[1-\mathds{1}_{\text{DC}}(x_{i+1},y_j) \right],\\
    T^{k}_{\text{U}}(x_i,y_j) &= \phi \left[\frac{C(x_i,y_{j+1})-C(x_i,y_j)}{4}\right]\left[1-\mathds{1}_{\text{DC}}(x_i,y_{j+1}) \right],\\
    T^{k}_{\text{D}}(x_i,y_j) &= \phi \left[\frac{C(x_i,y_{j-1})-C(x_i,y_j)}{4}\right]\left[1-\mathds{1}_{\text{DC}}(x_i,y_{j-1}) \right],\\
    T^{k}_{\text{S}}(x_i,y_j) &=  \left[1 - T^{k}_{\text{L}}(x_i,y_j) - T^{k}_{\text{R}}(x_i,y_j) - T^{k}_{\text{U}}(x_i,y_j) - T^{k}_{\text{D}}(x_i,y_j)\right].
\end{split}
\end{equation}
Finally, reflective boundaries are imposed to ensure T cells do not leave the domain. Figure \ref{fig:discrete_model_illustration} shows a typical example of allowable T cell movements, accounting for the boundary imposed by DCs.

\subsubsection*{T cell and dendritic cell interaction}
As detailed above, T cells are activated via stimulation from antigen-presenting DCs.
This occurs when T cells bind to DCs when they are in close proximity, which we determine to be one unit from any part of the dendritic cell.
To indicate when T cells are in this region, we define the indicator function, $\mathds{1}_{\text{A}}$.
Figure \ref{fig:discrete_model_illustration} shows an example of a T cell (the blue cell) that is in close proximity to the dendritic cell (in the grey region, where $\mathds{1}_{\text{A}}(x_m,y_n) = 1$), and therefore can accumulate stimulation due to the dendritic cell.
We prescribe a phenotype-discretisation, $\alpha \in \mathcal{R}_+$, and define $a_j=j \alpha$. At time-step $k$, a naïve T cell $i$ located at $(x_m,y_n)$, has an accumulated stimulation level of $A_i^k$. 
If a T cell is sufficiently close to a dendritic cell (indicated by $\mathds{1}_{\text{A}}(x_m,y_n) = 1$), then it will gain stimulation with probability:
\begin{align}\label{eq:stim_up}
    P^k_{+} = \psi_+  \left( 1 - \frac{A_i^k + \alpha}{A_{\text{max}}} \right) \mathds{1}_{\text{A}}(x_m,y_n),
\end{align}
with $A_i^{k+1} = \min \left\{A_i^k , A_{\text{max}} \right \}$, where $A_{\text{max}}$ is the maximum stimulation level required to become activated the T cell. {Here, $\psi_+$ represents the probability that a given T cell–DC interaction results in successful stimulation, and can be interpreted as a proxy for T cell receptor (TCR) and peptide major histocompatibility complex (pMHC) binding affinity. In reality, T cell affinity for a given antigen is not uniform across a population, but is governed by a Gaussian distribution reflecting the diversity of TCR sequences, whereby each TCR can bind to a range of peptides with varying affinities \cite{wooldridge2012single, sewell2012must}. Therefore, one approach to account for this heterogeneity is to allow $\psi_+$ to vary across T cells within the population. As this model is concerned with population-level activation dynamics, we take $\psi_+$ to represent the mean affinity across the population, which is the natural summary statistic of this distribution.}

Similarly, when a T cell is sufficiently far from the dendritic cell (indicated by $\mathds{1}_{\text{A}}(x_m,y_n) = 0$), it will lose stimulation with probability:
\begin{align}\label{eq:stim_down}
    P^k_{-} =  \psi_{-} \frac{A_i^{k} - \alpha}{A_{\text{max}}} \left[1-\mathds{1}_{\text{A}}(x_m,y_n) \right],
\end{align}
with $A_i^{k+1} = \max \left\{0, A_i^k \right \}$.
Importantly, once T cells reach a stimulation level of $A_{\text{max}}$, they are considered differentiated and no longer gain or lose stimulation.
For example, Figure \ref{fig:discrete_model_illustration} shows a T cell (yellow cell) that is sufficiently far from the dendritic cell (in the white region, where $\mathds{1}_{\text{A}}(x_m,y_n) = 0$) and therefore can lose stimulation.
% We note that both stimulation accumulation and loss are independent of one another, but vary in space. In this context, we define T cell sensitivity characteristics as follows: if a T cell has a small probability of stimulation uptake  ($\psi_{+} \ll 0.5$), then it is a low sensitive T cell. Similarly, if a T cell has a high probability of stimulation uptake ($\psi_{+} \gg 0.5$), then it is a high sensitive T cell. \textcolor{red}{This specification of T cell characteristics arise due  dictated by the actual differentiation in sensitivities that is found within the immune system \cite{}.}

\begin{figure}[h!]
    \centering
    \begin{subfigure}[b]{0.45\textwidth}
    \centering
    \includegraphics[width = \linewidth]{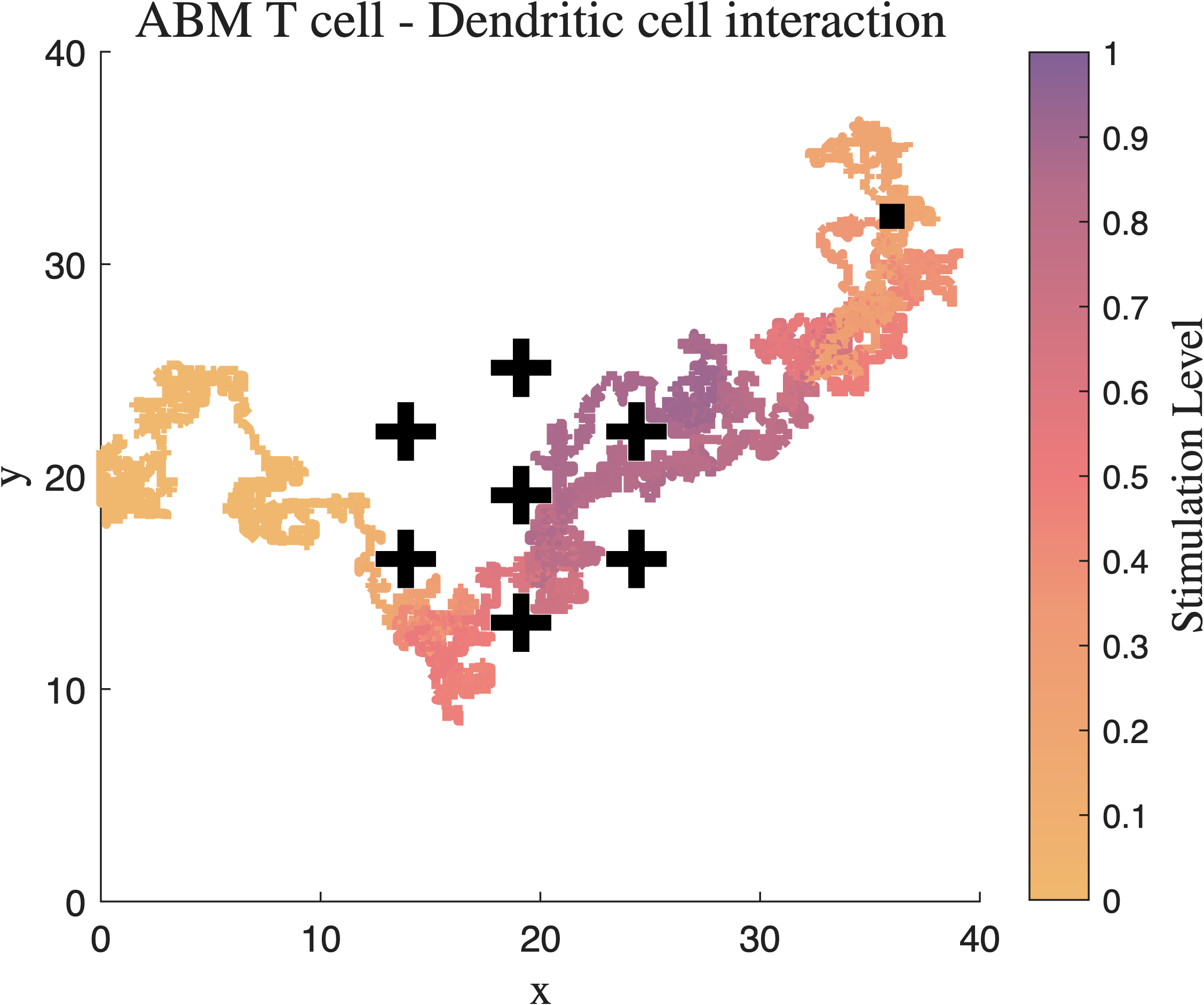}
    \caption{\label{fig:F0_walk}}
    \end{subfigure}
    \begin{subfigure}[b]{0.45\textwidth}
    \centering
    \includegraphics[width = \linewidth]{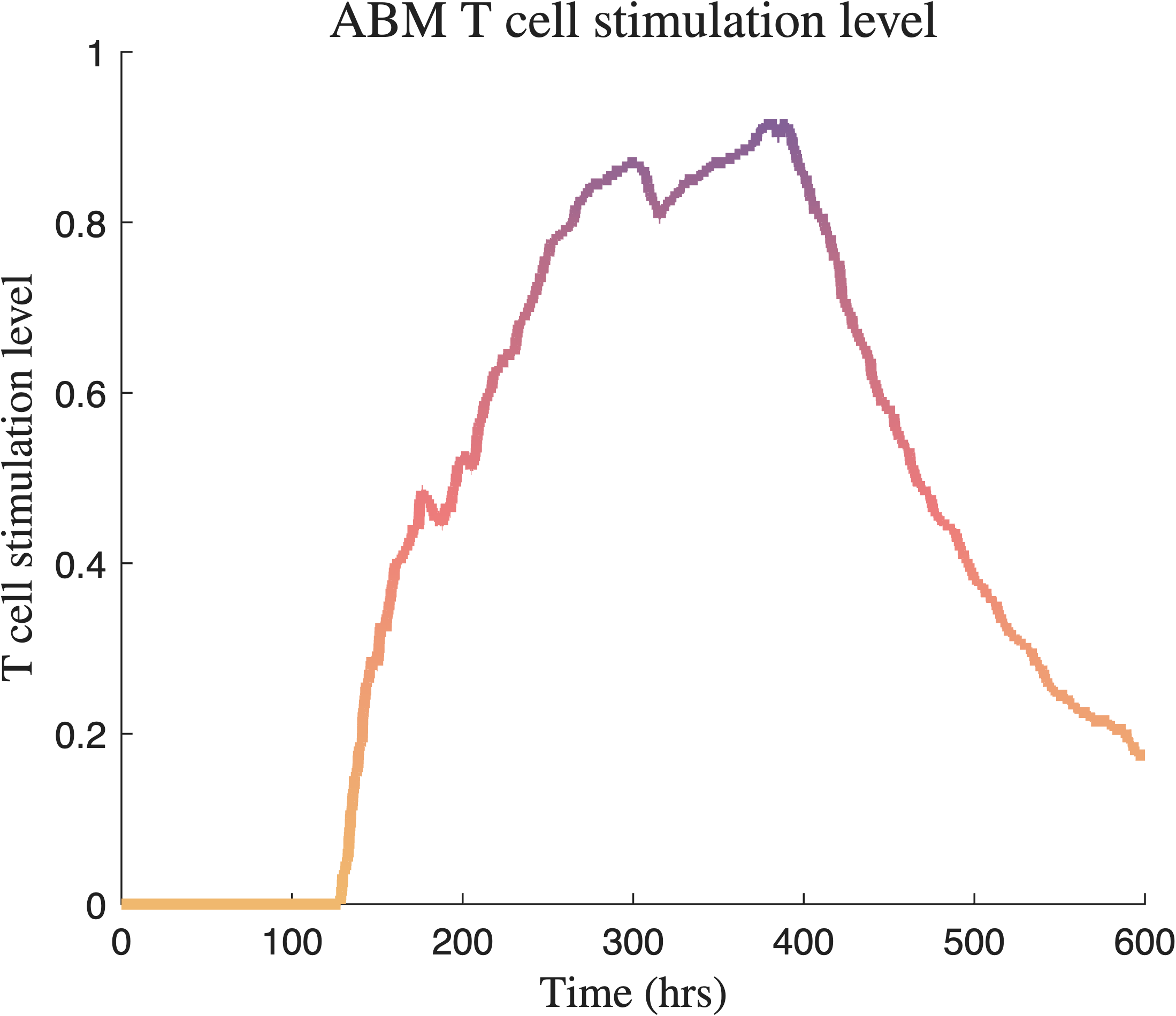}
    \caption{\label{fig:F0_Time}}
    \end{subfigure}
    
    \captionsetup{subrefformat=parens}
    \caption{\ref{fig:F0_walk} An example of a T cell biased random walk with 7 DCs in one cluster, with colour indicating the stimulation  level. Dark purple is high stimulation, bright yellow is no stimulation. \ref{fig:F0_Time} T cell stimulation level with time.}
    \label{fig:enter-label}
\end{figure}

\subsection{Continuum Model}
Here, we describe the phenotype-structured partial differential equation (PS-PDE) model. A detailed derivation is provided in the Supplementary Information, \ref{si:cont_derivation}.
In the continuum limit, allowing the time-step $\tau \rightarrow 0$, spatial-discretisation $\delta \rightarrow 0$, and phenotype-discretisation $\alpha \rightarrow 0$, we find that:
\begin{align}
    \frac{\theta \delta^2}{4 \tau} \rightarrow D \in \mathbb{R}_{+},  \quad \frac{\phi \delta^2}{4 \tau} \rightarrow \chi \in \mathbb{R}_{+},\quad \text{and} \quad \frac{\psi_{\pm} \alpha}{\tau} \rightarrow \mu_{\pm} \in \mathbb{R}_{+} \quad \text{as} \quad \tau,\delta,\alpha \rightarrow 0.
\end{align}
We obtain a structured partial-differential equation for the density of the T cell population, namely a phenotypic taxis-diffusion equation \cite{lorenzi2025phenotype}:
\begin{align} \label{eq:governing_pde}
    \frac{\partial u}{\partial t} & = \nabla \left\{ D \nabla u - u \, \chi \nabla C \right\} - \frac{\partial}{\partial a} \left\{ \left[ \mu_+\left(1-\frac{a}{A_{\text{max}}}\right) \mathds{1}_{\text{A}} -\mu_- \frac{a}{A_{\text{max}}} \left(1-\mathds{1}_{\text{A}}\right) \right] u\right\},
\end{align}
where $u(x,y,a,t)$ is the density of the T cell population with stimulation level $a$, $D$, the T cell diffusivity, $\chi$, the sensitivity of T cell chemotaxis, $\mu_+$, the rate at which T cells uptake stimulation due to DCs, $\mu_-$, the rate at which T cells lose stimulation, $A_{\text{max}}$, the maximum stimulation needed to become an activated T cell, and $C$, the chemokine gradient defined by Equation (\ref{eq:chemokine_field}).
To solve the above coupled PDE, the initial conditions we use are a constant T cell population at the left boundary, with no initial stimulation:
\begin{align}
    u(0,y,0,0) = u^{(0)}(y), \quad y \in \Omega, \quad \text{and} \quad u(x,y,a,0) = 0, \quad (x,y,a) \in \Omega_{x,a>0}, 
\end{align}
and no-flux boundary conditions:
\begin{equation}
\begin{split}
    \nabla u(x,y,a,t) \cdot \hat{n} &= 0, \quad (t,x,y,a) \in \mathbb{R}_{+} \times \partial\Omega,\\
    \nabla u(x,y,a,t) \cdot \hat{m} &= 0, \quad (t,x,y,a) \in \mathbb{R}_{+} \times \partial\Omega_{\text{DC}},\\
    \frac{\partial u}{\partial a} \Big\vert_{a=0} = \frac{\partial u}{\partial a} \Big\vert_{a=A_{\text{max}}} &= 0, \quad (t,x,y) \in \Omega,
\end{split}
\end{equation}

where $\hat{n}$ is the outward unit normal to the boundary $\partial\Omega$, and $\hat{m}$ is the outward unit normal to the boundary $\partial\Omega_{\text{DC}}$.

\subsubsection*{Steady-state analytical solution}
In general, we are not able to find a closed form solution to the system.
However, since we assume the chemokine gradient does not change, we are able to find a steady-state spatial distribution of the T cell population and the T cell stimulation level.
We first note that the spatial density of T cells is independent of $a$. Therefore, we can write the T cell density $N(x,y,t) = \int_0^{A_{\text{max}}} u(x,y,a,t) da$, which satisfies the taxis-diffusion PDE:
\begin{align*}
    \frac{\partial N}{\partial t} = \nabla \left\{ D \nabla N - N \chi \nabla C \right\},
\end{align*}
with boundary conditions:
\begin{align*}
        \nabla N \cdot \hat{n} &= 0, \quad (t,x,y) \in \mathbb{R}_{+} \times \partial\Omega,\\
    \nabla N \cdot \hat{m} &= 0, \quad (t,x,y) \in \mathbb{R}_{+} \times \partial\Omega_{\text{DC}},
\end{align*}
as before. We note that at steady state, since we impose a no-flux boundary condition, $N_{ss}$ satisfies $D\nabla N_{ss} - N_{ss} \chi \nabla C = 0$, which is a simple eigenvalue problem, with solution:
\begin{align}
    N_{ss}(x,y) = N_0 \exp \left[ \frac{\chi}{D} C(x,y) \right],
\end{align}
where $N_0$ is the normalisation constant.
Similarly, we describe the phenotypic density of the T cell stimulation at steady state, \mbox{$U(a) = U_A(a) + U_{A'}(a)$}. In this form, $U_A(a)$ describes the proportion of T cells with  stimulation level $a$, within the region of activation, and $U_{A'}(a)$ describes the proportion of T cells with stimulation level $a$ outside this region. From our governing PS-PDE, equation (\ref{eq:governing_pde}), at steady state, we write, assuming that $\mu_+,\mu_-,A_{\text{max}} > 0$:
\begin{align*}
    0 = \frac{d}{da} \left( -\left[ \mu_+\left(1-\frac{a}{A_{\text{max}}}\right) U_A \right] + \left[ \mu_- \frac{a}{A_{\text{max}}} U_{A'} \right] \right),
\end{align*}
and use the no-flux boundary at $a=0$ and $a=A_{\text{max}}$ to find the relationship between T cells inside and outside the activation region as:
\begin{align} \label{eq:a_balance}
 \mu_+\left(1-\frac{a}{A_{\text{max}}}\right) U_A  =  \mu_- \frac{a}{A_{\text{max}}} U_{A'} .
\end{align}
We now introduce a two-compartmental model for both $U_A$ and $U_{A'}$, which describes how the T cell stimulation changes in regions $A$ and $A'$:
% If we make the assumption that these two compartments are well mixed, we have the following equations:
\begin{equation}
    \begin{split}
        \frac{\partial U_A }{\partial t} &= - \frac{\partial}{\partial a}\left[ \mu_+\left(1-\frac{a}{A_{\text{max}}}\right) U_A \right] - \kappa_{A \rightarrow A'}U_A + \kappa_{A' \rightarrow A}U_{A'},\\
        \frac{\partial U_{A'} }{\partial t} &= + \frac{\partial}{\partial a}\left[ \mu_-\frac{a}{A_{\text{max}}} U_{A'} \right] + \kappa_{A \rightarrow A'}U_A - \kappa_{A' \rightarrow A}U_{A'}.
    \end{split}
\end{equation}

At steady state, the change in number of T cells within the region $A$ should balance those leaving and entering that region, i.e.:
\begin{align*}
    0 &= - \frac{d}{da}\left[ \mu_+\left(1-\frac{a}{A_{\text{max}}}\right) U_A \right] - \kappa_{A \rightarrow A'}U_A + \kappa_{A' \rightarrow A}U_{A'},
\end{align*}
where $\kappa_{A' \rightarrow A}$ and $\kappa_{A \rightarrow A'}$ are the rates of T cell exchange between regions $A'$ to $A$, and $A$ to $A'$, respectively. Using the relationship between $U_A$ and $U_{A'}$ from equation (\ref{eq:a_balance}), we find:
\begin{align*} 
    \frac{d}{da}\left[ \mu_+\left(1-\frac{a}{A_{\text{max}}}\right) U_A \right] + \kappa_{A \rightarrow A'}U_A - \kappa_{A' \rightarrow A} \frac{\mu_+}{\mu_-} \frac{A_{\text{max}}}{a}\left(1-\frac{a}{A_{\text{max}}}\right) U_A = 0.
\end{align*}
Rearranging and separating variables, we get the separable system:
\begin{align*}
    \int \frac{1}{U_A} d U_A = \int \left( \frac{1-\frac{\kappa_{A \rightarrow A'}A_{\text{max}}}{\mu_+}}{A_{\text{max}} - a} + \frac{\kappa_{A' \rightarrow A} A_{\text{max}}}{\mu_-}\frac{1}{a} \right) da,
\end{align*}
which has solution:
\begin{align*}
    \log(U_A) + \text{const.} = \left(\frac{\kappa_{A \rightarrow A'}A_{\text{max}}}{\mu_+} - 1\right) \log(A_{\text{max}} - a) + \frac{\kappa_{A' \rightarrow A} A_{\text{max}}}{\mu_-} \log(a).
\end{align*}\label{eq:U_A}
Taking the exponential of both sides, rewriting $\kappa_{1} = \frac{\kappa_{A' \rightarrow A} A_{\text{max}}}{\mu_-}$, and $\kappa_{2} =  \frac{\kappa_{A \rightarrow A'} A_{\text{max}}}{\mu_+}$ as dimensionless parameters, and requiring that the total distribution is unitary, we get:
\begin{align}
    U_A(a) = \bar{U} \,a \, f(a; \kappa_{1},\kappa_{2}) ,
\end{align}
and: 
\begin{align} \label{eq:U_Ap}
    U_{A'}(a) =  \bar{U} \frac{\mu_+}{\mu_-} \, (A_{\text{max}} - a) \, f(a; \kappa_{1},\kappa_{2}) ,
\end{align}
where $\bar{U} = \frac{1}{(A_{\text{max}})^{\kappa_1 + \kappa_2}} \left( \frac{\kappa_1 + \kappa_2}{\kappa_1 + \frac{\mu_+}{\mu_-} \kappa_2} \right)$, $f(a;\kappa_1,\kappa_2) = \frac{a^{\kappa_1-1}(A_{\text{max}}-a)^{\kappa_2-1}}{B(\kappa_1,\kappa_2)} $ is the beta distribution, and $B(\kappa_1,\kappa_2)$ the beta function, and:
\begin{align} \label{eq:analytic_approx_eq}
    U(a) = \bar{U}\left[ a + \frac{\mu_+}{\mu_-}
    \left(A_{\text{max}} - a \right) \right] f(a; \kappa_1, \kappa_2),
\end{align}
the steady-state analytic approximation of the T cell stimulation distribution.

We assume that the exchange of T cells between regions $A$ and $A'$ balances and is proportional to the mean number of T cells in the activation region of each DC. 
% If we let the number of DCs be $N_{DC}$, the number of dendritic cell clusters be $K_{DC}$, since we assume the DC interaction region to be unitary (for the general 2D case), we define $\kappa = \frac{ K_{DC}}{ N_{DC}}$.
We define the proportion of T cells in $A$ and $A'$ as $p_A$ and $p_{A'} = 1-p_A$, respectively, which can be directly computed from the steady-state function $N(x,y,t)$, given a geometry, as $p_A = \iint \mathds{1}_A N d \Omega $. We define $L_A$ as the characteristic length of the stimulation accumulation region ($A$) and similarly $L_{A'} = x_{\text{max}} - L_A$ (in 1D)  as the characteristic length  of the stimulation loss region ($A'$), and $\kappa$ as a scaling coefficient. We then write the dimensionless parameters:
\begin{align} \label{eq:kappas}
    \kappa_1 = \kappa \, p_A \frac{D}{L_A (x_{\text{max}} - L_A)} \frac{ A_{\text{max}}}{ \mu_-}, \quad \kappa_2 = \kappa \, (1-p_{A}) \frac{D}{L_A (x_{\text{max}} - L_A)} \frac{ A_{\text{max}}}{ \mu_+},
\end{align}
where $\frac{D}{L_A (x_{\text{max}} - L_A)}$ is the expected rate of a T cell in a given region to leave that region.
Biologically, we interpret $\kappa_1$ to be the characteristic scale of T cell stimulation loss, and $\kappa_2$ to be the characteristic scale of T cell stimulation uptake.
We can reduce the stimulation distribution of the T cells down to difference six cases:
\begin{enumerate}
    \item High stimulation regime: $\kappa_1 > \kappa_2$, and $\kappa_1 > 1$. 
    Unimodal T cell stimulation distribution, with a skew towards $a = A_{\text{max}}$.
    \item Balanced stimulation regime: $\kappa_1 \approx \kappa_2$, and $\kappa_1,\kappa_2 > 1$. Unimodal T cell stimulation distribution, about $a = A_{\text{max}}/2$.
    \item Low stimulation regime: $\kappa_2 > \kappa_1$, and $\kappa_2 > 1$. Unimodal T cell stimulation distribution, with a skew towards $a = 0$.
    \item Activation dominant regime, with coexistence: $\kappa_1 > \kappa_2$, and $\kappa_1 < 1$. Bimodal T cell stimulation distribution, weighted towards $a = A_{\text{max}}$.
    \item Coexistence regime: $\kappa_1 \approx \kappa_2$, and $\kappa_1,\kappa_2 < 1$. Bimodal T cell stimulation distribution, symmetric about $a = A_{\text{max}}/2$.
    \item Naïve dominant regime, with coexistence: $\kappa_2 > \kappa_1$, and $\kappa_2 < 1$. Bimodal T cell stimulation distribution, weighted towards $a = 0$.
\end{enumerate}
We also estimate the mean stimulation level as:
\begin{align}
    \mathbb{E}\left[ U(a) \right] = \frac{A_{\text{max}} \kappa_1}{\kappa_1 + \kappa_2 + 1} \left( \frac{1}{\kappa_1 + \frac{\mu_+}{\mu_-} \kappa_2} + 1 \right).
\end{align}
Lastly, since we know the mean stimulation level of T cells, we can also find the expected proportion of T cell activation:
\begin{align}
    \mathbb{E}\left[ P_A \right] = \frac{\mathbb{E}\left[ U(a) \right] }{ A_{\text{max}}}.
\end{align}

 \section{Results}

\subsection{ABM and PS-PDE Model Comparison} \label{sec:model_comparisions}
In this section, we provide a comparison between the PS-PDE and the ABM. We expect that as the number of T cells within the discrete model increases, and the spatial-discretisation, phenotype-discretisation and time-step approach those of the numerical discretisation used to compute the PS-PDE, the two models will converge to similar cell and stimulation distributions. Details on the numerical implementation of both the ABM and PS-PDE models can be found in the Supplementary Information \ref{si:numerics}.
We will show how each of the six profiles of T cell stimulation can be obtained depending on the size of the region of activation, and how the PS-PDE and ABM models produce comparable results. We will also show that these long term behaviours can be well predicted by the analytic approximation.

\subsubsection*{Simple one-dimensional domain}
\begin{figure}[H]
    \centering
    \includegraphics[width=0.3\linewidth]{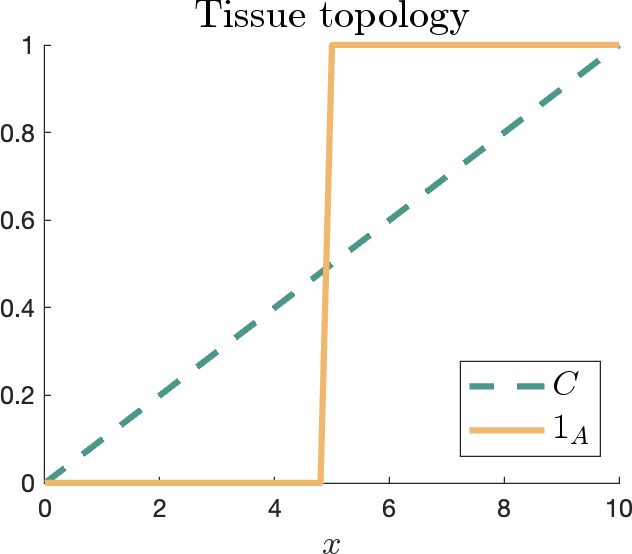}
    \caption{Schematic for the simple 1D tissue topology. T cells move up the gradient, $C$, (dashed green line), and become activated where $\mathds{1}_A = 1$ (solid yellow line).}
    \label{fig:1D_Topology}
\end{figure}

The first example we consider is a simple one-dimensional domain to best illustrate various stimulation distributions of the system. The tissue topology consists of a constant chemotaxis term towards the right of the domain, where T cells also gain stimulation. Figure \ref{fig:1D_Topology} shows an example with $\mathds{1}_A = 1$ for $x > 5$. Variations of this topology will include changing the domain at which T cells can uptake stimulation by changing the activation region, to resemble various Dendritic cell cluster sizes in higher dimensions.
We fix the diffusivity $D=0.5$, chemotactic sensitivity $\chi = 0.25$, the stimulation uptake rate $\mu_+ = 0.45$, and the stimulation degradation rate $\mu_- = 0.60$.
For the remainder of this section, we depict results from the discrete model by solid purple lines, results from the continuum model by dashed yellow lines, and the steady-state analytic approximation 
%(Equation (\ref{eq:analytic_approx_eq}))  
as a dotted blue line.
Since the movement dynamics are consistent across all Cases, we expect that the distributions of T cell position density are comparable. 
However, for Cases 1, 2 and 3, T cells have a higher stimulation capacity, $A_{\text{max}}$, and for Cases 4, 5 and 6, T cells have a lower stimulation capacity.
We will show that by simply varying the activation region, we can alter the characteristic scale of T cell stimulation loss, $\kappa_1$, and  the characteristic scale of T cell stimulation uptake, $\kappa_2$, and in turn alter the stimulation distribution of T cells.

\begin{figure}[h!]
    \centering
    \begin{subfigure}[b]{0.99\textwidth}
        \centering
        \includegraphics[width=1\linewidth]{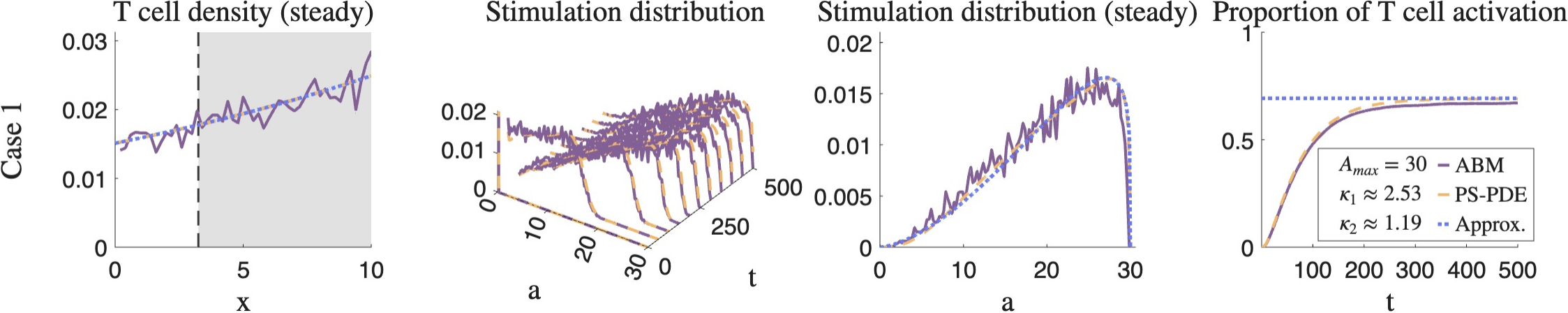}
        \caption{\label{fig:1D_case_1}}
    \end{subfigure}
    
    \begin{subfigure}[b]{0.99\textwidth}
        \centering
        \includegraphics[width=1\linewidth]{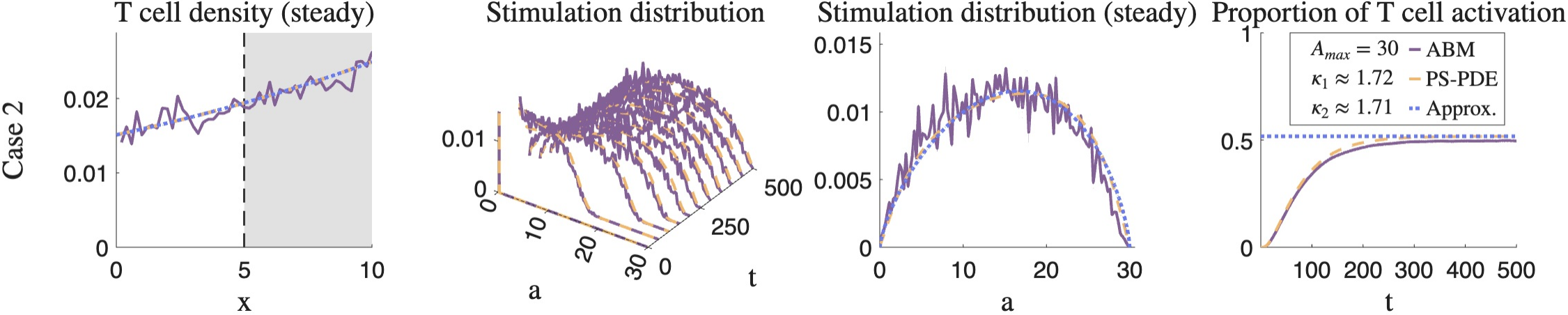}
        \caption{\label{fig:1D_case_2}}
    \end{subfigure}

    \begin{subfigure}[b]{0.99\textwidth}
        \centering
        \includegraphics[width=1\linewidth]{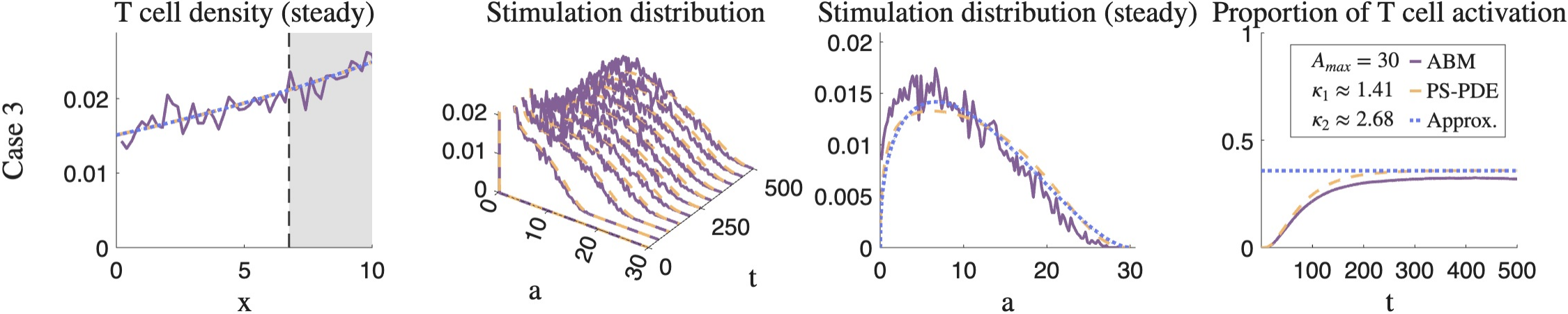}
        \caption{\label{fig:1D_case_3}}
    \end{subfigure}
    \caption{Case 1: skewed high T cell stimulation (top row), Case 2: balanced T cell stimulation (middle row), and Case 3: skewed low T cell stimulation (bottom row), for the ABM (purple), PS-PDE (yellow dashed) and analytic approximation (blue dotted) models. Left column shows the T cell density, with a large activation region (grey). Middle-left column shows the phenotype distribution in phenotype space ($a$), and how it changes in time. Middle-right column shows the long term steady-state distribution of the phenotype, with a skew to high T cell stimulation. Right column shows the proportion of T cells activation with time. We observe agreement across all models. Model parameters: $\chi=0.25$, $D=0.5$, $\mu_- = 0.45$, $\mu_+ = 0.6$, $A_{\text{max}} = 30$.}
    \label{fig:1D_cases_123}
    % Model parameters: $\chi=1$, $D=0.5$, $\mu_- = 0.4$, $\mu_+ = 0.6$, $A_{\text{max}} = 30$.

\end{figure}

\textbf{Case 1, High stimulation regime: $\kappa_1 > \kappa_2$, and $\kappa_1 > 1$}\\
Case 1 displays a high level stimulation distribution, with most T cells accumulating stimulation values towards $a=A_{\text{max}}$.
% Case 1 displays skewed high T cell activation, with the distribution in activation skewed towards $a=A_{\text{max}}$. 
In Figure \ref{fig:1D_case_1}, we observe how the stimulation distribution initially starts at zero, with all T cells being naïve, and then, gradually, most of the T cells gain stimulation, causing the distribution to skew towards $a=A_{\text{max}}$. However, we see that none of the T cells reach full activation in this example.

\begin{figure}[h!]
    \begin{subfigure}[b]{0.99\textwidth}
        \centering
        \includegraphics[width=1\linewidth]{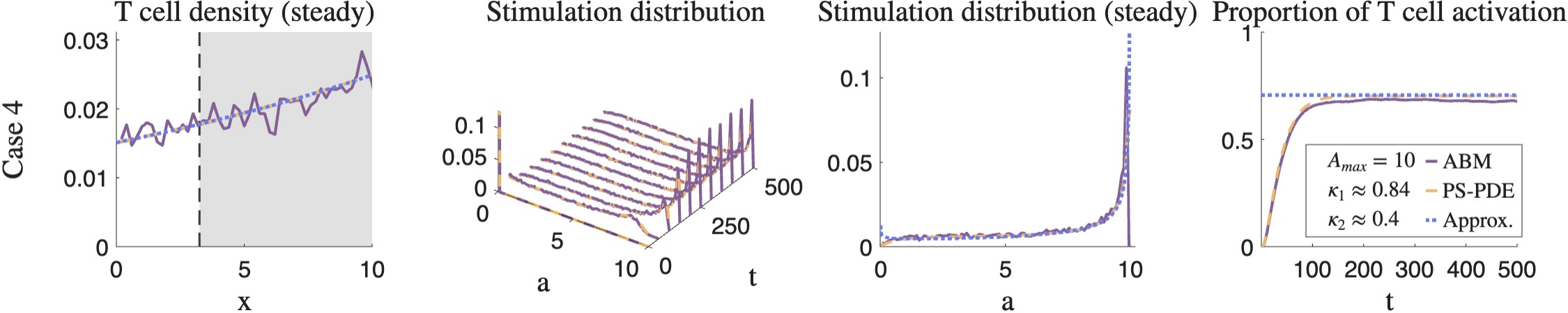}
        \caption{\label{fig:1D_case_4}}
    \end{subfigure}

    \begin{subfigure}[b]{0.99\textwidth}
        \centering
        \includegraphics[width=1\linewidth]{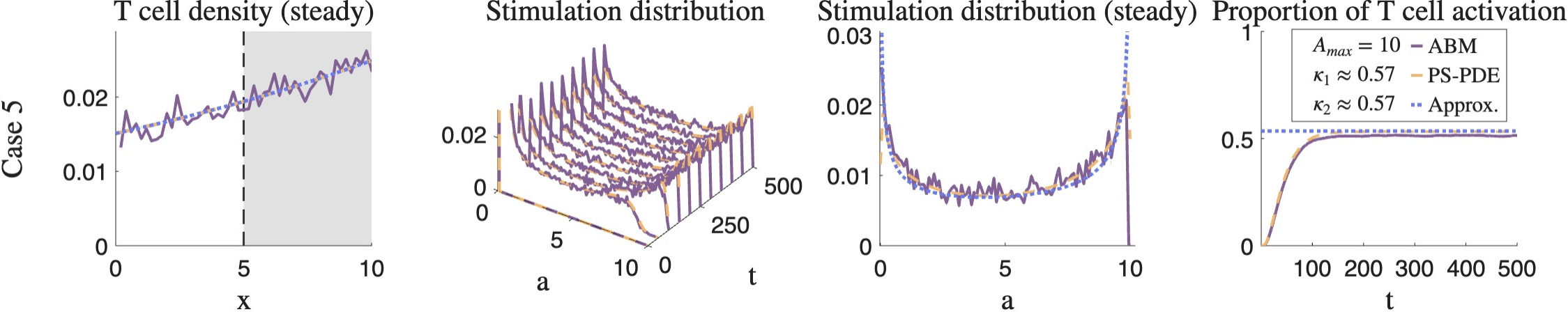}
        \caption{\label{fig:1D_case_5}}
    \end{subfigure}

    \begin{subfigure}[b]{0.99\textwidth}
        \centering
        \includegraphics[width=1\linewidth]{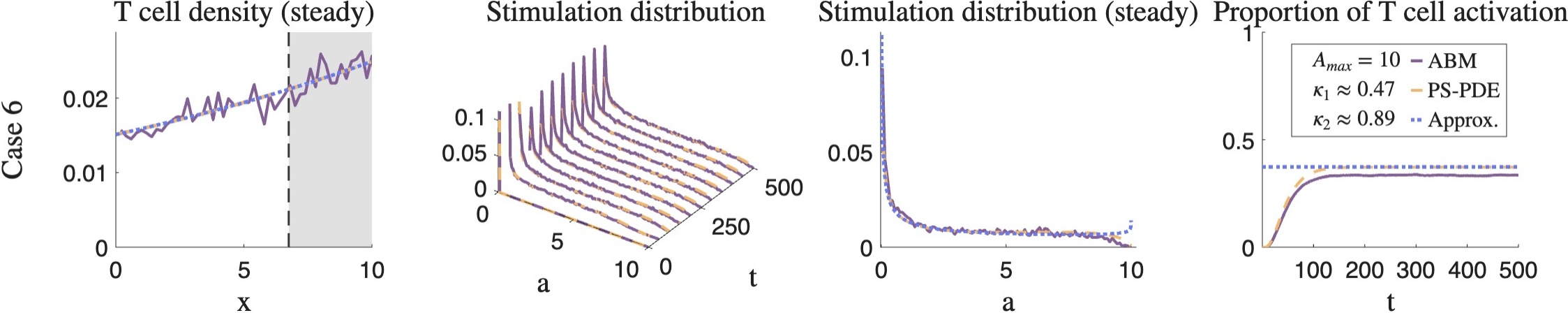}
        \caption{\label{fig:1D_case_6}}
    \end{subfigure}
    
    \caption{Case 4: Activation dominant (top row), Case 5: coexistence (middle row), and Case 6: naïve dominant (bottom row), for the ABM (purple), PS-PDE (yellow dashed) and analytic approximation (blue dotted) models. Left column shows the T cell density, with a large activation region (grey). Middle-left column shows the phenotype distribution in phenotype space ($a$), and how it changes in time. Middle-right column shows the long term steady-state distribution of the phenotype, with a skew to high T cell stimulation. Right column shows the proportion of T cells activation with time. We observe agreement across all models. Model parameters: $\chi=0.25$, $D=0.5$, $\mu_- = 0.45$, $\mu_+ = 0.60$, $A_{\text{max}} = 10$.}
    \label{fig:1D_cases_456}
    % Model parameters $\chi=1$, $D=0.5$, $\mu_- = 0.45$, $\mu_+ = 0.60$, $A_{\text{max}} = 10$.
\end{figure}

% \begin{figure}[H]
%     \centering
%     \includegraphics[width=0.98\linewidth]{Figures_v2/1D_case_1_v2.png}
%     \caption{Case 1: high T cell activation, for the ABM (purple), PS-PDE (yellow dashed) and analytic approximation (blue dotted) models. Left shows the T cell density, with a large activation region (grey). Middle-left shows the phenotype distribution in phenotype space ($a$), and how it changes in time. Middle-right shows the long term steady state distribution of the phenotype, with a skew to high T cell stimulation. Right shows the proportion of T cells activation with time. We observe agreement across all models.}
%     \label{fig:1D_case_1}
% \end{figure}

\textbf{Case 2, Balanced stimulation regime: $\kappa_1 \approx \kappa_2$, and $\kappa_1, \kappa_2 > 1$}\\
Case 2 (shown in Figure \ref{fig:1D_case_2}) displays a balanced stimulation distribution, with a mid-level, unimodal, T cell stimulation throughout the tissue, and the stimulation distribution of T cells centred around $a=A_{\text{max}/2}$. Again, in this example, none of the T cells have activated.
% \begin{figure}[H]
%     \centering
%     \includegraphics[width=0.98\linewidth]{Figures_v2/1D_case_3_v2.png}
%     \caption{Case 2: Mid level T cell activation, for the ABM (purple), PS-PDE (yellow dashed) and analytic approximation (blue dotted) models. Left shows the T cell density, with a large activation region (grey). Middle-left shows the phenotype distribution in phenotype space ($a$), and how it changes in time. Middle-right shows the long term steady state distribution of the phenotype, with a skew to high T cell stimulation. Right shows the proportion of T cells activation with time. We observe agreement across all models.}
%     \label{fig:1D_case_3}
% \end{figure}

\textbf{Case 3, Low stimulation regime: $\kappa_2 > \kappa_1$, and $\kappa_2 > 1$}\\
Case 3 displays a low level stimulation distribution, with most T cells accumulating stimulation values towards $a=0$.
% Case 3 displays skewed low T cell activation, with the distribution skewed towards $a=0$.
In Figure \ref{fig:1D_case_3}, we see how the stimulation distribution increases, but not to a large level, remaining low. While none of the T cells become activated in this instance, we also observe how none of the T cells have a stimulation level of zero in the PS-PDE model. However, in the ABM, we do find T cells with zero stimulation.
% \begin{figure}[H]
%     \centering
%     \includegraphics[width=0.98\linewidth]{Figures_v2/1D_case_2_v2.png}
%     \caption{\textcolor{red}{Todo}}
%     \label{fig:1D_case_2}
% \end{figure}

\textbf{Case 4, Activation dominant regime,: $\kappa_1 > \kappa_2$, and $\kappa_1 < 1$}\\
Case 4 displays an example of a highly activated level of T cells, with a prominent peak around $a=A_{\text{max}}$, see Figure \ref{fig:1D_case_4}. We see that a large proportion of T cells have become fully activated, with a mismatch between the analytical approximation, and the ABM/PS-PDE models in the tail of the stimulation distribution, around $a=0$. The analytic approximation predicts an additional peak in the stimulation distribution around $a=0$, but, this additional peak is not observed in the ABM and PS-PDE models.
% \begin{figure}[H]
%     \centering
%     \includegraphics[width=0.98\linewidth]{Figures_v2/1D_case_4_v2.png}
%     \caption{\textcolor{red}{Todo}}
%     \label{fig:1D_case_4}
% \end{figure}

\textbf{Case 5, Coexistence regime: $\kappa_1 \approx \kappa_2$, and $\kappa_1, \kappa_2 < 1$}\\
Case 5 displays an example of a coexistence regime of both highly activated and naïve T cells, with a symmetric, bi-modal T cell stimulation distribution, see Figure \ref{fig:1D_case_5}. In this case, we see prominent peaks at both $a=0$ and $a=A_{\text{max}}$, showing that T cells are present in both the fully naïve and also fully activated forms. Again, in this case, we also observe a mismatch between the analytical approximation, and the ABM/PS-PDE models, now in both tails of the stimulation distribution, around $a=0$, and $a=A_{\text{max}}$.
% \begin{figure}[H]
%     \centering
%     \includegraphics[width=0.98\linewidth]{Figures_v2/1D_case_6_v2.png}
%     \caption{\textcolor{red}{Todo}}
%     \label{fig:1D_case_6}
% \end{figure}

\textbf{Case 6, Naïve dominant regime, with coexistence: $\kappa_2 > \kappa_1$, and $\kappa_2 < 1$}\\
Case 6 displays an example of a predominantly naïve T cell population, with a prominent peak around $a=0$, see Figure \ref{fig:1D_case_6}. This shows us that most T cells are completely naïve, with no accumulated stimulation. In this case, we again observe a mismatch between the analytical approximation, and the ABM/PS-PDE models, in the tail of the stimulation distribution, around $a=A_{\text{max}}$. Similarly to case 4, the analytic approximation predicts an additional peak in the stimulation distribution around $a=A_{\text{max}}$, which is not observed in the ABM and PS-PDE models.
% \begin{figure}[H]
%     \centering
%     \includegraphics[width=0.98\linewidth]{Figures_v2/1D_case_5_v2.png}
%     \caption{\textcolor{red}{Todo}}
%     \label{fig:1D_case_5}
% \end{figure}

For all of the 1D examples shown here, we can see that the distribution of the accumulated stimulation can be controlled by the parameters $\kappa_1$ and $\kappa_2$. Importantly, we see that, by simply varying the region where T cells can accumulate stimulation, we can drastically alter the stimulation distribution of T cells. 
We also observe that for all the examples presented, the ABM and the PS-PDE model result in comparable findings for T cell position density, stimulation distribution and also proportion of activation. Our analytic approximation, which utilises the PS-PDE model at steady state, is capable of predicting the T cell stimulation distribution at steady state.

\subsubsection*{A single Dendritic cell activation site in two-dimensions}
\begin{figure}[H]
    \centering
    \includegraphics[width=1\linewidth]{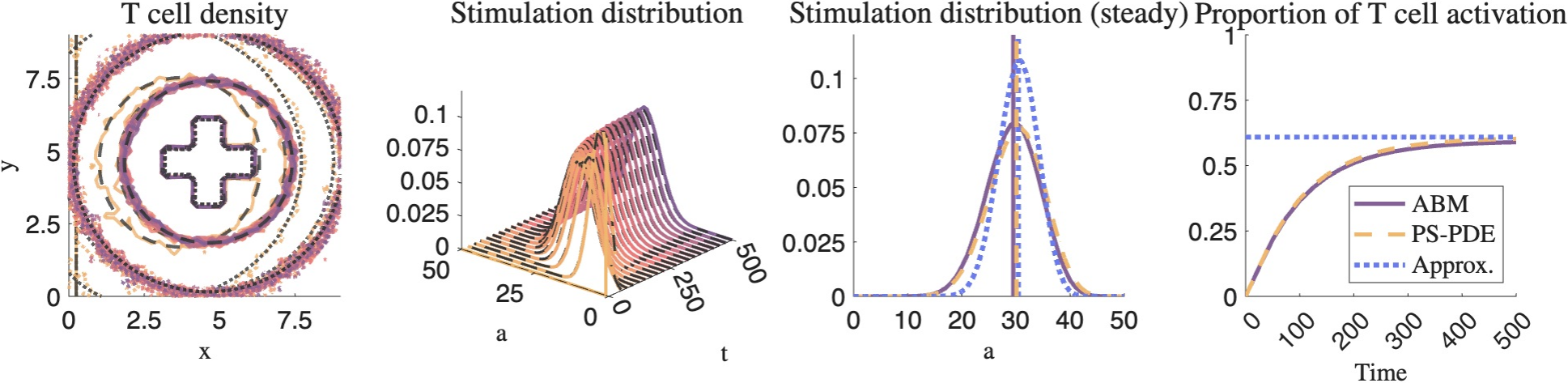}
    \caption{Comparison between PS-PDE model (dashed) and ABM (solid) for a single dendritic cell activation site. Left column shows the T cell density with T cell density contours of $u_1 = 1 \times 10^{-3}$ (dashed) and $u_2 = 5\times10^{-4}$ (dotted). Middle left column shows the stimulation distribution for the PS-PDE model (dashed black) and the ABM (solid coloured from yellow/light to purple/dark to indicate time), with time. Middle right column shows the steady-state stimulation distribution for the ABM (solid purple), PS-PDE model  dashed yellow) and the analytic approximation (dotted blue), and corresponding mean values (vertical corresponding lines). Right column shows the proportion of T cell activation over time for the ABM (solid purple), PS-PDE model (dashed yellow) and the analytic approximation (dotted blue). Model parameters: $\chi=1$, $D=0.5$, $\mu_- = 0.5$, $\mu_+ = 0.5$, and $A_{\text{max}} = 50$.}
    \label{fig:2D_activation_decay}
\end{figure}

We now consider a single dendritic cell, in a small section of the tissue. In this manner, we again compare trajectories from the discrete model ABM and the continuum PS-PDE model, while fixing the dendritic cell topology. For the T cell density, we present contour plots, that show how the fixed contour line progresses with time, with light (yellow) solid lines showing initial times, and dark (purple) solid lines showing final times. For the T cell stimulation distribution, we present tri-dimensional plots (phenotypic state and intensity, changing in time) for both the ABM (solid, with light/yellow lines showing initial times, and dark/purple lines showing final times) and PS-PDE models (dashed).  Results are given in Figure \ref{fig:2D_activation_decay}.

Considering the T cell density (Figure \ref{fig:2D_activation_decay} left), we see that the contour levels $u_1=1 \times 10^{-3}$ and $u_2= 5\times 10^{-4}$ of both models closely agree for all times shown. 
Considering the stimulation distribution of the T cells  (Figure \ref{fig:2D_activation_decay} middle-left), we observe that the T cells gain stimulation in a bulk, with the distribution starting off initially low, and then moving to higher stimulation. None of the T cells become fully activated, but none are fully naïve either. We see that both the ABM (coloured solid lines) and the PS-PDE model (dashed black lines) align well, with both models resulting in the same distributions at all the times presented here. 
When comparing the steady-state stimulation distribution of the ABM (solid purple), the PS-PDE model (dashed yellow), and the analytic approximation (dotted blue, Figure \ref{fig:2D_activation_decay} middle-right), we can see that the mean stimulation value of all three models align well, the shapes and skewness of the distributions coincide, but the height and spread of the analytic approximation does not exactly match the other two, simulated models. 
As a long term summary statistic, the proportion of T cell activation (Figure \ref{fig:2D_activation_decay} right) shows good agreement with all three descriptions.

The above analysis is for a single set of parameters only. However, we are interested in understanding the sensitivity of our framework when varying the 
% T cell sensitivity to chemotaxis ($\chi$), T cell diffusivity ($D$),
 maximum stimulation level of T cells required to become activated, $A_{\text{max}}$,
the rate of antigen loss ($\mu_-$) and also the rate of antigen accumulation ($\mu+$). These model parameters define the characteristics of the T cells:
to understand the effect of variations, we perform a parameter sensitivity analysis in $\mu-$ and $\mu+$, for a fixed set of parameters $\chi$ and $D$, and for a range of $A_{\text{max}}$.

\begin{figure}[H]
    \centering
    \includegraphics[width=0.99\linewidth]{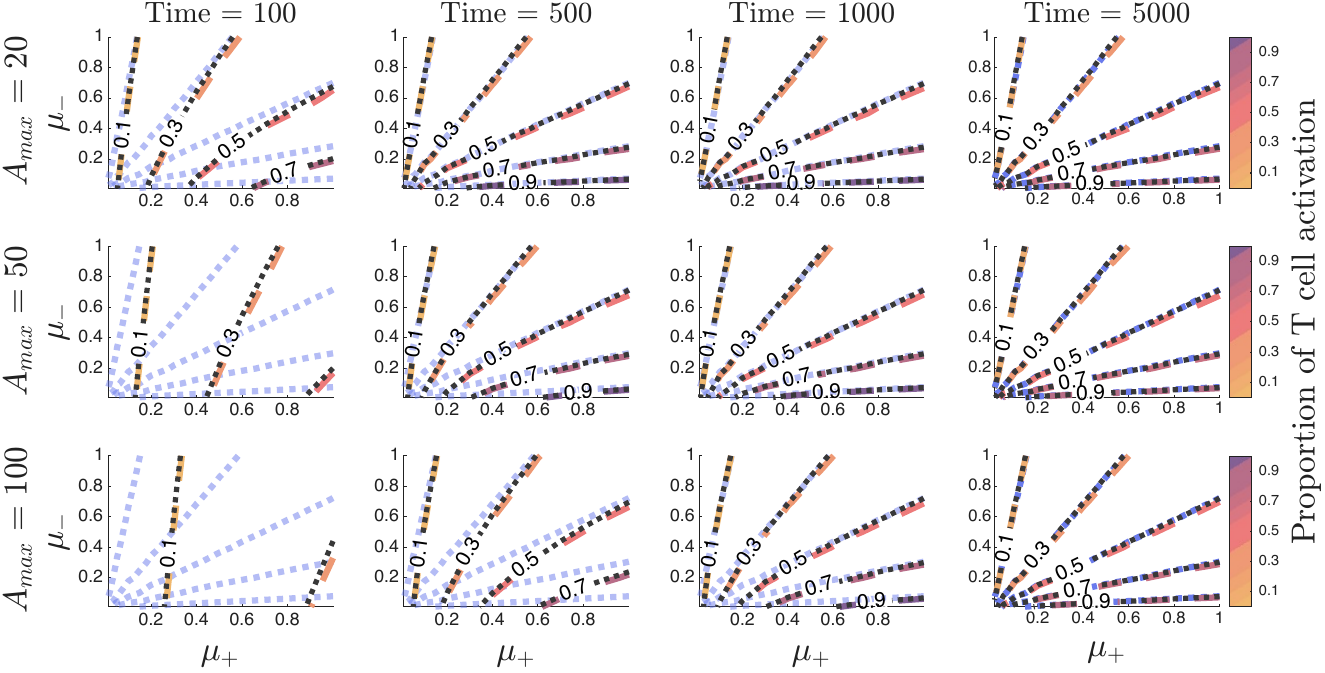}
    \caption{Comparison in proportion of T cell activation between PS-PDE (dashed black), ABM (dashed coloured), and the approximation (dotted blue), for fixed values of $A_{max}=20$ (top row), $A_{max}=20$ (middle row), and $A_{max}=100$ (bottom row), at times $t=100$ (first column), $t=500$ (second column), $t=1000$ (third column), and $t=5000$ (fourth column). We see that all results match the approximation at the final times, but exhibit different timescales arriving at the steady state approximation. Namely, the higher $A_{\text{max}}$ takes longer to reach equivalent proportion of T cell activation levels.}
    \label{fig:2D_Comparision}
\end{figure}
% From Figure \ref{fig:2D_Comparision}, we observe that, for low chemotactic sensitivity, $\chi$, the presented contours for each model are well aligned. We observe that as $\chi$, the models stay well aligned for most regions in the $\mu_-$--$\mu_+$ space, however differing slightly in the top right corner. We can therefore conclude that the T cell activation dynamics present in the two dimensional, single dendritic cell activation ABM example is captured well by the PS-PDE approximation.

Figure \ref{fig:2D_Comparision} presents the parameter sensitivity analysis of the PS-PDE (dashed black), ABM (dashed coloured), and the approximation (dotted blue), for different $A_{\text{max}}$, calculated at four different time points.
For a low level of required stimulation to become activated, $A_{\text{max}}$, the proportion of T cell activation rapidly approaches the steady state value.
We also see that, as we increase $A_{\text{max}}$, the time required to reach this steady state also increases.
Note also that, given sufficient time, both the ABM and the PS-PDE converge to the steady state approximation. We can therefore see that, for the model parameters presented here, there is a high degree of agreement between both the ABM and PS-PDE models.
These results also show two distinct regions in parameter space where the proportion of T cell activation transitions from being dependant on $\mu_+$ (namely the upper left region) to being to dependant on $\mu_-$ (namely the lower right region). In these regions, small changes in the dependant parameter result in a higher increase in the proportion of T cell activation.
Further details that consider multiple DCs in a cluster (i.e., and hence consider an increase in the activation region) in two-dimensions are provided in the Supplementary Information \ref{si:multiple_DCs}.

\subsection{DC clustering alters T cell stimulation distribution} \label{sec:main_results}

Here, we present the key results of our study, considering a variety of T cell types, based on their characteristics: i) how readily a T cell binds, and  interacts with DCs to gain stimulation ($\mu_+$), ii) how readily T cells losse stimulation ($\mu_-$). To do so, we first parametrise our model based on known values, see Table \ref{tab:parameter_tabs}.

\begin{table}[H]
    \centering
    \begin{tabular}{|c|c|c|c|}
    \hline
    \textbf{Parameter} & \textbf{Physical meaning} & \textbf{Value} & \textbf{Reference} \\
    \hline
    $D$     &  T cell diffusivity               & $50 \, \mu m^2 \text{min}^{-1}$ & \cite{mempel2004t,banigan2015heterogeneous,celli2012many} \\
    $\chi$  &  T cell DC chemotaxis sensitivity & $100 \, \mu m^2 \text{min}^{-1}$ & \cite{stachowiak2006homeostatic}\\
    $\mu_+$ &  T cell stimulation uptake            & $[0.01, 0.1, 1]$ & --\\
    $\mu_-$ &  T cell stimulation decay            & $[0.01, 0.1, 1]$ & --\\
    $A_{\text{max}}$ & Maximum stimulation capacity &  $100$ & --\\
    $\#$ DCs & Number of DCs / DC density & 128 / $\sim 3\%$ &\cite{johnson2021lymph,czepielewski2023resident,beltman2007lymph} \\
    $\delta$ & Spatial discretisation & $0.25$ & -- \\
    $\tau$ & Time discretisation & $0.025$ & -- \\
    $\alpha$ & Stimulation discretisation & $0.2$ & -- \\
    \hline
    \end{tabular}
    \caption{Parameter values used to perform numerical simulation and sensitivity analyses.}
    \label{tab:parameter_tabs}
\end{table}

\begin{figure}[h!]
    % \centering
    % \includegraphics[width=0.95\linewidth]{Figures_v2/MainResultsFig_Full.png}
    \centering
    \begin{subfigure}[b]{0.32\textwidth}
        \centering \includegraphics[height = 5cm]{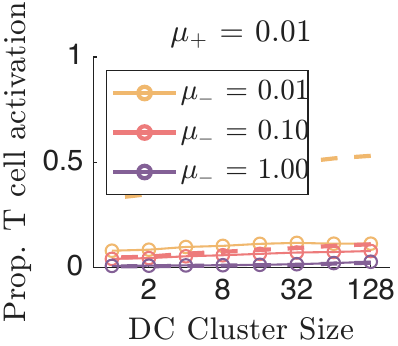}
        \caption{\label{fig:MainResultsFig_Full_subplot_1}}
    \end{subfigure}
    \begin{subfigure}[b]{0.32\textwidth}
        \centering \includegraphics[height = 5cm]{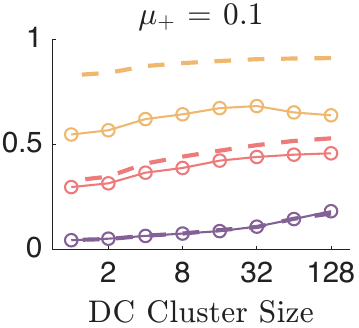}
        \caption{\label{fig:MainResultsFig_Full_subplot_2}}
    \end{subfigure}
    \begin{subfigure}[b]{0.32\textwidth}
        \centering \includegraphics[height = 5cm]{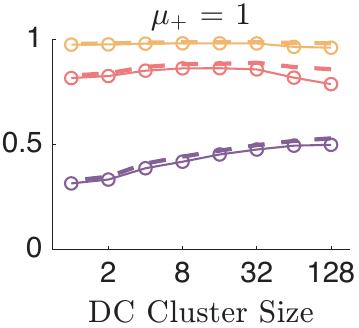}
        \caption{\label{fig:MainResultsFig_Full_subplot_3}}
    \end{subfigure}

    \centering
    \begin{subfigure}[b]{0.32\textwidth}
        \centering \includegraphics[height = 5cm]{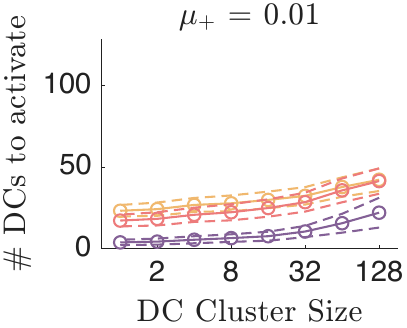}
        \caption{\label{fig:MainResultsFig_Full_subplot_4}}
    \end{subfigure}
    \begin{subfigure}[b]{0.32\textwidth}
        \centering \includegraphics[height = 5cm]{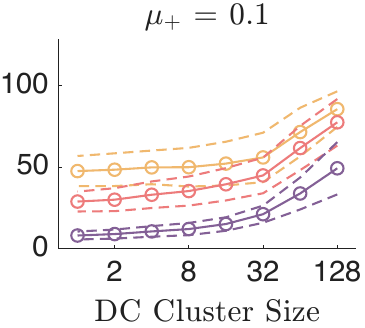}
        \caption{\label{fig:MainResultsFig_Full_subplot_5}}
    \end{subfigure}
    \begin{subfigure}[b]{0.32\textwidth}
        \centering \includegraphics[height = 5cm]{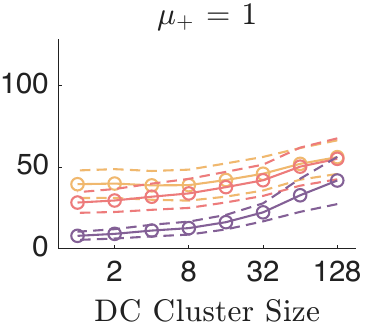}
        \caption{\label{fig:MainResultsFig_Full_subplot_6}}
    \end{subfigure}
    
    \caption{Sensitivity analysis on the spatial clustering of DCs, for various T cell characteristics. Top row shows the mean proportion of T cell activation over 10 unique Dendritic cell topologies after 48 hours, comparing results from the ABM (solid line) and the analytic approximation (dashed line). Bottom row shows the mean number of DCs required for T cells to become activated (solid line) with the $95\%$ confidence interval (dashed line). 
    Results are shown for a low rate of stimulation uptake (\ref{fig:MainResultsFig_Full_subplot_1} and \ref{fig:MainResultsFig_Full_subplot_4} respectively), mid-rate of stimulation uptake (\ref{fig:MainResultsFig_Full_subplot_2} and \ref{fig:MainResultsFig_Full_subplot_5} respectively), and high rate of stimulation uptake (\ref{fig:MainResultsFig_Full_subplot_3} and \ref{fig:MainResultsFig_Full_subplot_6} respectively), for low rate of stimulation decay (yellow), mid-rate of stimulation decay (red), and high rate of stimulation decay (purple).}
    \label{fig:MainResultsFig_Full}
\end{figure}

We utilise the discrete ABM to determine how the system evolves in time, and show how DC clustering can affect the level of T cell activation. Figure \ref{fig:MainResultsFig_Full} shows the level of T cell activation for various ($\mu_-,\mu_+$) pairs, and how DC clustering influences this level. 
The analytic approximation, representing the long-term behaviour, is shown as a dashed line, while the ABM results at the final time $t=48$ hrs are shown as a solid line (see Figures \ref{fig:MainResultsFig_Full_subplot_1} - \ref{fig:MainResultsFig_Full_subplot_3}). We also present the average number of unique DCs that the T cell population interacts with: this measure provides an understanding of how heterogeneously the T cell population is activated. The associated envelope around the mean, which we chose to be one standard deviation from the mean (see Figures \ref{fig:MainResultsFig_Full_subplot_4} - \ref{fig:MainResultsFig_Full_subplot_6}), is also shown.

Not all scenarios reach their long term analytic approximation in the prescribed time, as the time scales for each scenario changes.
Therefore, for T cells that have a lower rate of stimulation uptake ($\mu_+$), it is expected these require a longer time to reach their final dynamics.
This phenomenon is particularly emphasised when we look at how the T cell stimulation distribution changes in time (see Figure \ref{fig:MainResultsFig_TCellPrifile}).

From Figures \ref{fig:MainResultsFig_Full_subplot_4} - \ref{fig:MainResultsFig_Full_subplot_6}, for all of the T cell characteristics we consider, DC clustering enhances the mean heterogeneity of T cell activation, indicating that T cells become activated from more DCs when DCs are clustered.

\subsubsection*{DC clustering enhances mean T cell activation, and affects timescale of T cell activation}
From Figures \ref{fig:MainResultsFig_Full_subplot_1} - \ref{fig:MainResultsFig_Full_subplot_3}, we see that for the majority of T cell characteristics considered, DC clustering enhances the mean T cell activation.

\begin{figure}[h!]
    % \centering
    % \includegraphics[width=0.95\linewidth]{Figures_v2/MainResultsFig_TCellPrifile.png}
    \centering
    \begin{subfigure}[b]{0.32\textwidth}
        \centering \includegraphics[height = 3.5cm]{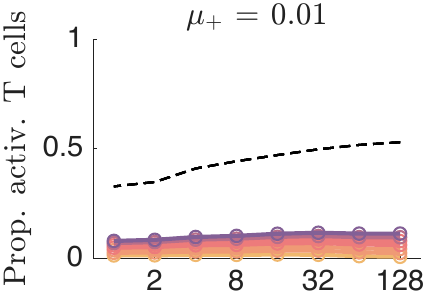}
        \caption{\label{fig:MainResultsFig_TCellPrifile_subplot_1}}
    \end{subfigure}
    \begin{subfigure}[b]{0.32\textwidth}
        \centering \includegraphics[height = 3.5cm]{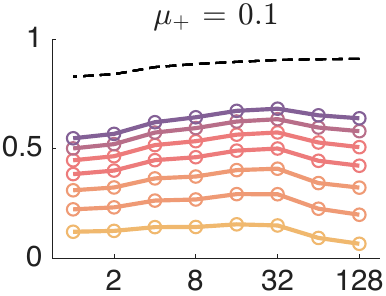}
        \caption{\label{fig:MainResultsFig_TCellPrifile_subplot_2}}
    \end{subfigure}
    \begin{subfigure}[b]{0.32\textwidth}
        \centering \includegraphics[height = 3.5cm]{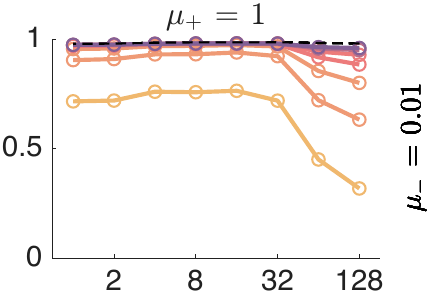}
        \caption{\label{fig:MainResultsFig_TCellPrifile_subplot_3}}
    \end{subfigure}

    \centering
    \begin{subfigure}[b]{0.32\textwidth}
        \centering \includegraphics[height = 3.5cm]{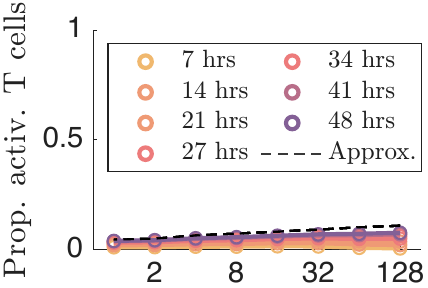}
        \caption{\label{fig:MainResultsFig_TCellPrifile_subplot_4}}
    \end{subfigure}
    \begin{subfigure}[b]{0.32\textwidth}
        \centering \includegraphics[height = 3.5cm]{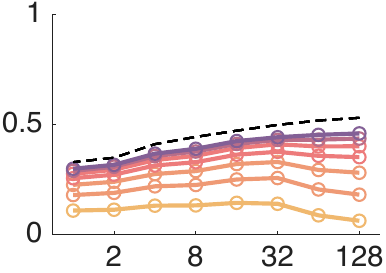}
        \caption{\label{fig:MainResultsFig_TCellPrifile_subplot_5}}
    \end{subfigure}
    \begin{subfigure}[b]{0.32\textwidth}
        \centering \includegraphics[height = 3.5cm]{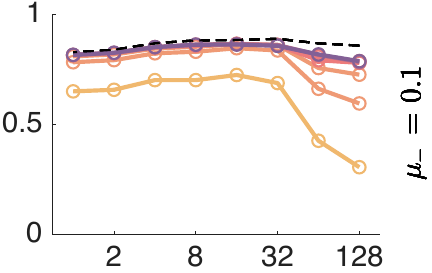}
        \caption{\label{fig:MainResultsFig_TCellPrifile_subplot_6}}
    \end{subfigure}

    \centering
    \begin{subfigure}[b]{0.32\textwidth}
        \centering \includegraphics[height = 3.8cm]{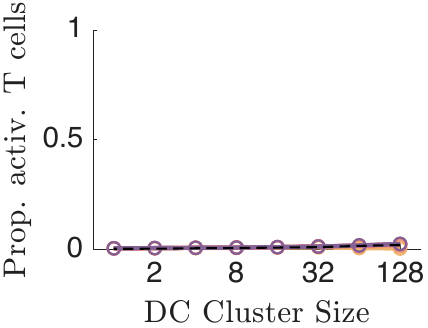}
        \caption{\label{fig:MainResultsFig_TCellPrifile_subplot_7}}
    \end{subfigure}
    \begin{subfigure}[b]{0.32\textwidth}
        \centering \includegraphics[height = 3.8cm]{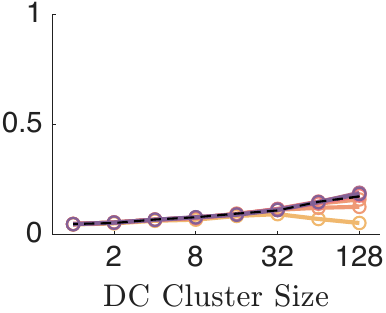}
        \caption{\label{fig:MainResultsFig_TCellPrifile_subplot_8}}
    \end{subfigure}
    \begin{subfigure}[b]{0.32\textwidth}
        \centering \includegraphics[height = 3.8cm]{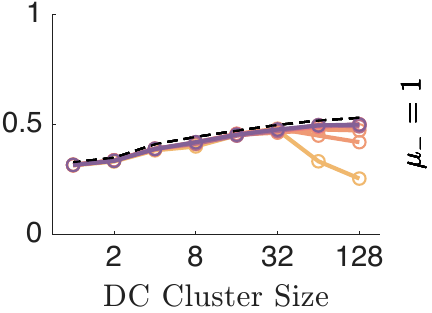}
        \caption{\label{fig:MainResultsFig_TCellPrifile_subplot_9}}
    \end{subfigure}
    
    \caption{Sensitivity of the proportion of activated T cells to spatial clustering of DCs. Solid, coloured lines show results from the the ABM, and the dashed black line shows the analytic approximation. Times shown are 7 hours (yellow/light), 14, 21, 27, 34, 41, and 48 hours (purple/dark). T cell characteristics vary by plot, with left (top) column (row) being low rate of stimulation uptake (decay), middle column (row) being mid-level stimulation uptake (decay), and right (bottom) column (row) being high rate of stimulation uptake (decay).}
    \label{fig:MainResultsFig_TCellPrifile}
\end{figure}

By considering the proportion of activated T cells, we see that T cells with a lower rate of stimulation uptake, $\mu_+$, result in a lower proportion of activated T cells (see Figures \ref{fig:MainResultsFig_TCellPrifile_subplot_1}, \ref{fig:MainResultsFig_TCellPrifile_subplot_4}, and \ref{fig:MainResultsFig_TCellPrifile_subplot_7}). Similarly, those with a higher rate of stimulation decay, $\mu_-$, also exhibit a lower proportion of activated T cells (see Figures \ref{fig:MainResultsFig_TCellPrifile_subplot_7} - \ref{fig:MainResultsFig_TCellPrifile_subplot_9}). Intuitively, as $\mu_+$ increases, so does the proportion of activated T cells, as expected. Turning to the influence of DC clustering on the proportion of activation, we observe that, for high levels of stimulation decay ($\mu_- = 1$ , Figures \ref{fig:MainResultsFig_TCellPrifile_subplot_7} - \ref{fig:MainResultsFig_TCellPrifile_subplot_9}), DC clustering promotes T cell activation. For a medium stimulation decay rate ($\mu_- = 0.1$, Figures \ref{fig:MainResultsFig_TCellPrifile_subplot_4} - \ref{fig:MainResultsFig_TCellPrifile_subplot_6}), this promoting effect begins to taper off as $\mu_+$ increases, and even reverses for high levels of DC clustering (see Figure \ref{fig:MainResultsFig_TCellPrifile_subplot_6}). Finally, for a low stimulation decay rate ($\mu_- = 0.01$ Figures \ref{fig:MainResultsFig_TCellPrifile_subplot_1} - \ref{fig:MainResultsFig_TCellPrifile_subplot_3}), the proportion of activated T cells does not reach the approximated steady state value within the permitted time for both low and medium levels of stimulation uptake.

% To discuss the timescale of T cell activation, if we consider the rightmost column of Figure \ref{fig:MainResultsFig_TCellPrifile}, (i.e. $\mu_+ = 1$, Figures \ref{fig:MainResultsFig_TCellPrifile_subplot_3}, \ref{fig:MainResultsFig_TCellPrifile_subplot_6}, and \ref{fig:MainResultsFig_TCellPrifile_subplot_9}), we see that scenarios with high levels of DC cluster require more time to reach high levels of T cell activation compared to cells of a similar characteristic, but with lower DC clustering.
% If we compare this observation to the middle column of Figure \ref{fig:MainResultsFig_TCellPrifile}, (i.e. $\mu_+ = 0.1$, Figures \ref{fig:MainResultsFig_TCellPrifile_subplot_2}, \ref{fig:MainResultsFig_TCellPrifile_subplot_5}, and \ref{fig:MainResultsFig_TCellPrifile_subplot_8}), we see that all T cells populations activate on a comparable timescale, as all of the T cell activation profiles exhibit consistent increases, irrespective of DC clustering.

Considering the timescale of T cell activation, for a high rate of stimulation uptake, $\mu_+ = 1$, Figures \ref{fig:MainResultsFig_TCellPrifile_subplot_3}, \ref{fig:MainResultsFig_TCellPrifile_subplot_6}, and \ref{fig:MainResultsFig_TCellPrifile_subplot_9}), scenarios with high levels of DC clustering require more time to reach high levels of T cell activation compared to those with lower DC clustering, but otherwise similar T cell characteristics. 
In contrast, for a medium level of stimulation uptake, $\mu_+ = 0.1$, Figures \ref{fig:MainResultsFig_TCellPrifile_subplot_2}, \ref{fig:MainResultsFig_TCellPrifile_subplot_5}, and \ref{fig:MainResultsFig_TCellPrifile_subplot_8}), all T cell populations become activated on a comparable timescale, with their proportion of activated T cell levels exhibiting a consistent increase, irrespectively of DC clustering.

From these results, we identify two key effects of DC clustering: firstly, DC clustering enhances the mean level of T cell activation, and secondly, it reduces the timescale of T cell activation for those with medium and high stimulation uptake rates.

\subsubsection*{DC clustering enhances heterogeneous T cells activation}

\begin{figure}[h!]
    
    \centering
    \begin{subfigure}[b]{0.32\textwidth}
        \centering \includegraphics[height = 3.6cm]{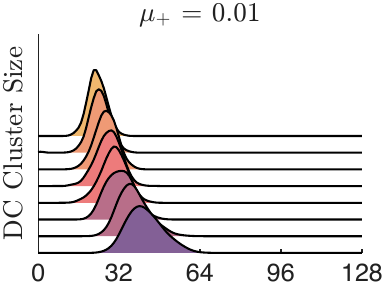}
        \caption{\label{fig:MainResultsFig_TCellHeterogeneity_subplot_1}}
    \end{subfigure}
    \begin{subfigure}[b]{0.32\textwidth}
        \centering \includegraphics[height = 3.6cm]{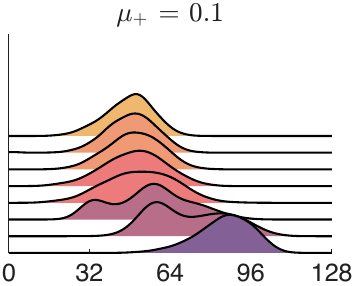}
        \caption{\label{fig:MainResultsFig_TCellHeterogeneity_subplot_2}}
    \end{subfigure}
    \begin{subfigure}[b]{0.32\textwidth}
        \centering \includegraphics[height = 3.6cm]{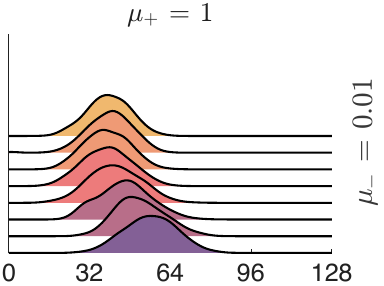}
        \caption{\label{fig:MainResultsFig_TCellHeterogeneity_subplot_3}}
    \end{subfigure}

    \centering
    \begin{subfigure}[b]{0.32\textwidth}
        \centering \includegraphics[height = 3.8cm]{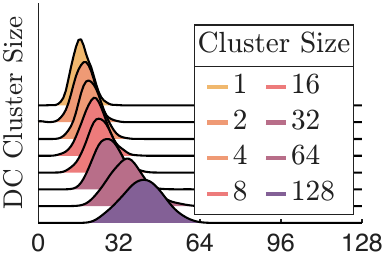}
        \caption{\label{fig:MainResultsFig_TCellHeterogeneity_subplot_4}}
    \end{subfigure}
    \begin{subfigure}[b]{0.32\textwidth}
        \centering \includegraphics[height = 3.2cm]{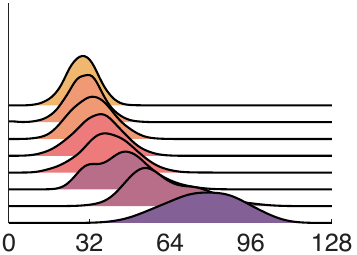}
        \caption{\label{fig:MainResultsFig_TCellHeterogeneity_subplot_5}}
    \end{subfigure}
    \begin{subfigure}[b]{0.32\textwidth}
        \centering \includegraphics[height = 3.2cm]{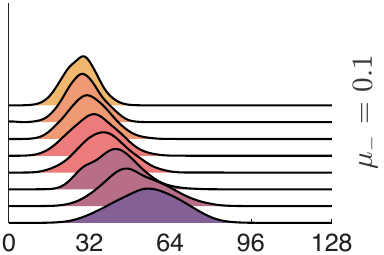}
        \caption{\label{fig:MainResultsFig_TCellHeterogeneity_subplot_6}}
    \end{subfigure}

    \centering
    \begin{subfigure}[b]{0.32\textwidth}
        \centering \includegraphics[height = 3.7cm]{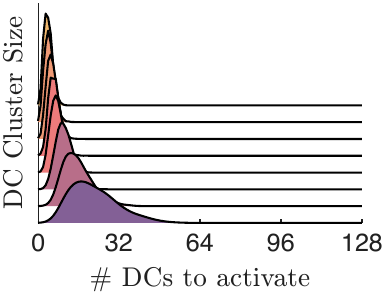}
        \caption{\label{fig:MainResultsFig_TCellHeterogeneity_subplot_7}}
    \end{subfigure}
    \begin{subfigure}[b]{0.32\textwidth}
        \centering \includegraphics[height = 3.7cm]{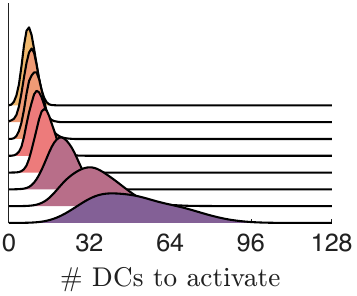}
        \caption{\label{fig:MainResultsFig_TCellHeterogeneity_subplot_8}}
    \end{subfigure}
    \begin{subfigure}[b]{0.32\textwidth}
        \centering \includegraphics[height = 3.75cm]{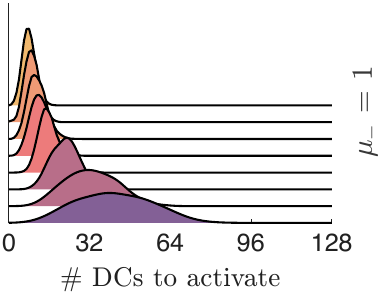}
        \caption{\label{fig:MainResultsFig_TCellHeterogeneity_subplot_9}}
    \end{subfigure}

    \caption{Sensitivity of the heterogeneity of T cell activation to spatial clustering of DCs, after 48 hours. Distributions shown for each T cell characteristic are for various Dendritic cell cluster size from low clustering (yellow, 1 DC per cluster), to high clustering (purple, 128 DCs per cluster). 
    T cell characteristics vary by plot, with left (top) column (row) being low rate of stimulation uptake (decay), middle column (row) being mid-level stimulation uptake (decay), and right (bottom) column (row) being high rate of stimulation uptake (decay).}
    \label{fig:MainResultsFig_TCellHeterogeneity}
\end{figure}

While both the ABM and PS-PDE descriptions are equally capable of describing how the T cell population becomes activated,
% by describing the stimulation distribution within the population, 
only the ABM description can identify how the T cells are actually engaging with the DCs, and more precisely, with which DCs.
In the ABM, for each T cell in our population, we can identify which DCs they engage with, and moreover, how much stimulation they acquired from that particular DC. 
We therefore measure the heterogeneity of T cell activation, by quantifying how many DCs a T cell engages with. We then produces a distribution of how many DCs the T cell population has had interactions with. To present summary statistics, we fit a kernel density from this distribution, with results shown in Figure \ref{fig:MainResultsFig_TCellHeterogeneity} for various T cell characteristics.

We categorise T cell activation as heterogeneous if the T cell population interact with a large number of DCs, and homogeneous if they interact with only a small number of DCs.
As seen in Figure \ref{fig:MainResultsFig_TCellHeterogeneity}, for all T cell characteristics considered, a higher level of DC clustering consistently results in more heterogeneous T cell activation.
However, for T cells with a low stimulation uptake rate ($\mu_+ = 0.01$), this increase in heterogeneity is relatively small (Figures \ref{fig:MainResultsFig_TCellHeterogeneity_subplot_1}, \ref{fig:MainResultsFig_TCellHeterogeneity_subplot_4}, and \ref{fig:MainResultsFig_TCellHeterogeneity_subplot_7}). Similarly, for T cells with a high stimulation uptake rate ($\mu_+ = 1$), while we see an enhancement in the heterogeneity of T cell activation with DC clustering, the improvement is not significant (Figures \ref{fig:MainResultsFig_TCellHeterogeneity_subplot_3}, \ref{fig:MainResultsFig_TCellHeterogeneity_subplot_6}, and \ref{fig:MainResultsFig_TCellHeterogeneity_subplot_9}).

In contrast, for a medium level of stimulation uptake ($\mu_+ = 0.1$), a high level of DC clustering strongly promotes heterogeneous T cell activation across all three stimulation decay rates considered (Figures \ref{fig:MainResultsFig_TCellHeterogeneity_subplot_2}, \ref{fig:MainResultsFig_TCellHeterogeneity_subplot_5}, and \ref{fig:MainResultsFig_TCellHeterogeneity_subplot_8}), regardless of whether the stimulation decay rate is low, medium, or high.

From these results, we conclude that DC clustering enhances the heterogeneity of T cell activation, most prominently for T cells with an intermediate stimulation uptake rate, that is, those that neither uptake stimulation too readily nor too slowly.

\section{Discussion}

We have presented a discrete, agent-based model (ABM) of T cells interacting with dendritic cells (DCs), where T cells undergo a biased random walk, chemotactically migrating towards the DC population.
% The DCs are described by static stencilled agents that do not move, but secrete a chemokine to attract the migrating T cells.
The DCs are described by static stencilled agents that secrete a chemokine to attract the migrating T cells, and we determined a systematic method for quantifying and varying DC cluster sizes.
T cells interact with DCs when they are in close proximity to the DC, accumulating stimulation, and hence activation signal.
As T cells lose contact with the DC, their accumulated stimulation decays, resulting in a loss in activation signal.

We performed a continuum limit approximation, to derive the continuum phenotypically structured partial differential equation (PS-PDE) that describes the T cell density via a taxis-diffusion equation.
The phenotypic structure within the PDE is capable of describing how the T cells interact with the DCs, via an advection component in the phenotypic dimension.

Using our PS-PDE, we performed a steady-state analysis, noting that the phenotypic advection does not influence cell motion, to identify the T cell spatial density. Using this, we were able to derive an approximation of the stimulation profile, given by Equations (\ref{eq:U_A}) and (\ref{eq:U_Ap}).
This steady-state approximation utilises the fact that the exchange rate of T cells in and out of a DC cluster is proportional to the T cells diffusion, the number of cells in the region, and the size of the activation regions.
Under these assumptions, the stimulation profile of the T cell population is then characterised by the parameters $\kappa_1$ and $\kappa_2$, given in Equation (\ref{eq:kappas}), which we interpret to be the characteristic scales of T cell stimulation loss and uptake, respectively.

In Section \ref{sec:model_comparisions}, we presented simulation results of both the ABM, PS-PDE, and the analytical approximation, showing that both the ABM and PS-PDE qualitatively match, and moreover that the analytical approximation of the stimulation distribution matched both the shape and mean of both the ABM and PS-PDE, for both the 1D and 2D problems.
We show that the timescale of T cell activation is influenced by T cell characteristics, such as the amount of accumulated stimulation needed to elicit a response, the rate of stimulation uptake, and the rate of stimulation decay.

Finally, in Section \ref{sec:main_results} we present a sensitivity analysis for multiple different T cell features.
We show that T cell characteristics dictate the stimulation distribution of T cells (Figures \ref{fig:MainResultsFig_Full_subplot_1} - \ref{fig:MainResultsFig_Full_subplot_3}), and also the timescale at which the T cells become activated (Figure \ref{fig:MainResultsFig_TCellPrifile}). Our results show how activation is influenced by DC clustering, but DC clustering does not appear to be a key driver of T cell activation level, since we see that DC clustering only marginally improves activation level across most regimes considered.

However, our results do demonstrate that the heterogeneity of T cell activation is influenced by DC clustering (Figures \ref{fig:MainResultsFig_Full_subplot_4} - \ref{fig:MainResultsFig_Full_subplot_6}).
We found that under particular T cell characteristics, T cells can either be activated in a highly heterogeneous manner or in a homogeneous manner (Figure \ref{fig:MainResultsFig_TCellHeterogeneity}).
We therefore hypothesis that dendritic cell clustering is advantageous for those T cells that exhibit the appropriate characteristics to take advantage of this heterogeneous activation.
From this, we further hypothesis DC clustering to selectively promote T cell activation of particular characteristics, which could provide context towards the serial engagement model of T cell activation \cite{valitutti2012serial}.

Naturally, our model description contains many necessary biological limitations.
For example, T cell crowding \cite{bogle2012lattice, munoz2014t}, antigen specificity \cite{ philip2022cd8} and T cell egress \cite{steele2023t} are not accounted for by our model, but are known to contribute significantly to their activation dynamics. {An important extension of this model would be to allow stimulation update parameters to vary across the T cell population. This would enable the model to capture how affinity heterogeneity within a population influences activation dynamics, for instance, whether a population with a broad distribution of affinities activates more robustly or more slowly than one concentrated near the mean \cite{wooldridge2012single, sewell2012must}. Such an extension would be particularly relevant in the context of tumour-infiltrating T cell populations, where TCR affinity distributions may differ markedly from healthy cohorts.}
Moreover, we model the lymph node topology as a simple, unobstructed domain, whereas lymph nodes are known to contain scaffold that assist T cell trafficking, know as the fibroblastic reticular cell network \cite{novkovic2016topological}.
We have also only described the stimulation uptake process by a simple probabilistic (or rate based) mechanism, rather than the highly complex T cell-dendritic cell binding mechanism that is know to initiate T cell activation \cite{guermonprez2002antigen, bousso2008t, steinman2003tolerogenic}. Furthermore, we do not consider motile DCs, variability in DC shape, or a time-varying chemokine gradient, each of which may influence T cell activation dynamics. We note that exploring these model limitations would be a fruitful source for future work.

Nevertheless, while a complete description of T cell and dendritic cell interaction would provide a more realistic insight into the mechanism of T cell activation, in this work we have striven to only consider the minimal necessary dynamics that govern the process, to allow a clear understanding of how DC clustering may influence T cell activation.

The results presented in this work may further provide insight into T cell activation mechanics.
One possible clinical implication of this work is the identification of T cell properties that may benefit from DC clustering, which could aid experimentalists in developing therapeutics that target these specific characteristics, resulting in improved patient outcomes.

% Possible outcomes and implications of this work include clinically identifying those T cell characteristics that may potentially benefit from DC clustering.
% By doing so, this work can aid experimentalists to develop therapeutics  that target these specific T cell characteristics, resulting in improved patient outcomes.

\section*{Data and Code Availability}
The manuscript has associated code and data in a repository available at \url{https://github.com/DGermano8/ABMToPSPDEImmuneModel.git}.  

\section*{Author Contributions}
% D.P.J.G, F.F, R.P.A, P.P.L, and P.S.K conceptualised research;\\
% D.P.J.G  designed research;\\
% D.P.J.G  performed research;\\
% D.P.J.G contributed software tools;\\
% D.P.J.G analysed data;\\
% F.F, R.P.A, and P.S.K supervised research;\\
% D.P.J.G wrote the paper;\\
% D.P.J.G, F.F, R.P.A,  P.P.L, and P.S.K reviewed the paper;\\
% F.F, R.P.A,  P.P.L, and P.S.K acquired funding.
Conceptualisation: Domenic P.J. Germano, Federico Frascoli, Robyn P. Araujo, Peter P. Lee, Peter S. Kim.\\
Formal analysis: Domenic P.J. Germano.\\
Funding acquisition: Federico Frascoli, Robyn P. Araujo, Peter P. Lee, Peter S. Kim.\\
Investigation: Domenic P.J. Germano.\\
Methodology: Domenic P.J. Germano.\\
Project administration: Domenic P.J. Germano.\\
Software: Domenic P.J. Germano.\\
Supervision: Federico Frascoli, Robyn P. Araujo, Peter S. Kim.\\
Validation: Domenic P.J. Germano.\\
Visualisation: Domenic P.J. Germano.\\
Writing – original draft: Domenic P.J. Germano.\\
Writing – review and editing: Domenic P.J. Germano, Federico Frascoli, Robyn P. Araujo, Peter P. Lee, Peter S. Kim.

\section*{Acknowledgements}
The authors gratefully acknowledges support from the Australian Research Council Discovery Project (DP230100485).

% Bibliography
\printbibliography[heading=bibintoc]
\newpage

\renewcommand\thesection{SI \arabic{section}}
\renewcommand\thesubsection{SI\arabic{section} \arabic{subsection}}

% Reset section and subsection counters
\setcounter{section}{0}
\setcounter{subsection}{0}
% \resetlinenumber
% \setcounter{page}{1}
\title{Supplementary Information: Spatial dynamic modelling to understand how dendritic cell clustering affects T cell activation}
\maketitle
\section{Continuum model derivation} \label{si:cont_derivation}
To derive the continuum limit of the discrete model, we employ the principle of mass conservation, noting that $\mathds{1}(x_i)$ dictates the appropriate no-flux boundary conditions:
% \begin{align} \label{eq:master_equation}
%     \begin{split}
%     u_{i,j}^{k+1} = u_{i,j}^{k} &+ \frac{\theta}{2}\left(u_{i-1,j}^{k} + u_{i+1,j}^{k} \right) \left[1- \mathds{1}(x_{i}) \right]  \\
%     & - \frac{\theta}{2}\left(\left[1- \mathds{1}(x_{i-1}) \right] + \left[1- \mathds{1}(x_{i+1}) \right]\right) u_{i,j}^{k} \\
%     & + \frac{\phi}{2}\left(\left[ c(x_{i}) - c(x_{i-1}) \right]_+ u_{i-1,j}^{k} + \left[ c(x_{i}) - c(x_{i+1}) \right]_+ u_{i+1,j}^{k} \right) \left[1- \mathds{1}(x_{i}) \right] \\
%     & - \frac{\phi}{2}\left( \left[c(x_{i-1}) - c(x_i)\right]_+\left[1- \mathds{1}(x_{i-1}) \right] + \left[c(x_{i+1}) - c(x_i)\right]_+\left[1- \mathds{1}(x_{i+1}) \right]  \right) u_{i,j}^{k} \\
%     & + \psi_+ \left( 1 - \frac{a_{j-1}}{A_{\text{max}}} \right) \mathds{1}_A(x_i) u_{i,j-1}^{k} - \psi_+ \left( 1 - \frac{a_{j}}{A_{\text{max}}} \right) \mathds{1}_A(x_i) u_{i,j}^{k} \\
%     & + \psi_- \left(\frac{a_{j+1}}{A_{\text{max}}} \right) \mathds{1}_A(x_i) u_{i,j+1}^{k} - \psi_- \left(\frac{a_{j}}{A_{\text{max}}} \right) \mathds{1}_A(x_i) u_{i,j}^{k}.
%     \end{split}
% \end{align}
\begin{align} \label{eq:master_equation}
    \begin{split}
    u_{i,j}^{k+1} = u_{i,j}^{k} &+ \frac{\theta}{2}\left(u_{i-1,j}^{k} + u_{i+1,j}^{k} \right) - \frac{\theta}{2}\left( u_{i,j}^{k} + u_{i,j}^{k} \right) \\
    & + \frac{\phi}{2}\left(\left[ c(x_{i}) - c(x_{i-1}) \right]_+ u_{i-1,j}^{k} + \left[ c(x_{i}) - c(x_{i+1}) \right]_+ u_{i+1,j}^{k} \right)  \\
    & - \frac{\phi}{2}\left( \left[c(x_{i-1}) - c(x_i)\right]_+  + \left[c(x_{i+1}) - c(x_i)\right]_+ \right) u_{i,j}^{k} \\
    & + \psi_+ \left( 1 - \frac{a_{j-1}}{A_{\text{max}}} \right) \mathds{1}_A(x_i) u_{i,j-1}^{k} - \psi_+ \left( 1 - \frac{a_{j}}{A_{\text{max}}} \right) \mathds{1}_A(x_i) u_{i,j}^{k} \\
    & + \psi_- \left(\frac{a_{j+1}}{A_{\text{max}}} \right) \left[1-\mathds{1}_A(x_i)\right] u_{i,j+1}^{k} - \psi_- \left(\frac{a_{j}}{A_{\text{max}}} \right) \left[1-\mathds{1}_A(x_i)\right] u_{i,j}^{k}.
    \end{split}
\end{align}
We can utilise the relations, which hold when $\tau$, $\delta$ and $\alpha$ are sufficiently small:
\begin{gather*}
t_k \approx t, \quad t_{k \pm 1} \approx t \pm \tau, \quad x_{i} \approx x, \quad x_{i \pm 1} \approx x \pm \delta, \quad a_{j} \approx a, \quad a_{j \pm 1} \approx a \pm \alpha,\\
    u_{i,j}^{k} \approx u(t,x,a), \quad u_{i,j}^{k\pm 1} \approx u(t \pm \tau,x,a), \quad u_{i\pm 1,j}^{k} \approx u(t ,x \pm \delta,a), \quad \quad u_{i,j\pm 1}^{k} \approx u(t ,x,a \pm \alpha),\\
    c(x_i) \approx c(x), \quad c(x_{i \pm 1}) = c(x \pm \delta).
\end{gather*}
We can then rewrite (\ref{eq:master_equation}) as:
% \begin{align}
%     \begin{split}
%     u(t+\tau,x,a) = u(t,x,a) &+ \frac{\theta}{2}\left(u(t,x-\delta,a) + u(t,x+\delta,a) \right) \left[1- \mathds{1}(x) \right]  \\
%     & - \frac{\theta}{2}\left(\left[1- \mathds{1}(x-\delta) \right] + \left[1- \mathds{1}(x+\delta) \right]\right) u(t,x,a) \\
%     & + \frac{\phi}{2}\left(\left[ c(x) - c(x-\delta) \right]_+ u(t,x-\delta,a) + \left[ c(x) - c(x+\delta) \right]_+ u(t,x+\delta,a) \right) \left[1- \mathds{1}(x) \right] \\
%     & - \frac{\phi}{2}\left( \left[c(x-\delta) - c(x)\right]_+\left[1- \mathds{1}(x-\delta) \right] + \left[c(x+\delta) - c(x)\right]_+\left[1- \mathds{1}(x+\delta) \right]  \right) u(t,x,a) \\
%     & + \psi_+ \left( 1 - \frac{a-\alpha}{A_{\text{max}}} \right) \mathds{1}_A(x) u(t,x,a-\alpha) - \psi_+ \left( 1 - \frac{a}{A_{\text{max}}} \right) \mathds{1}_A(x) u(t,x,a) \\
%     & + \psi_- \left(\frac{a+\alpha}{A_{\text{max}}} \right) \mathds{1}_A(x) u(t,x,a+\alpha) - \psi_- \left(\frac{a}{A_{\text{max}}} \right) \mathds{1}_A(x) u(t,x,a).
%     \end{split}
% \end{align}
\begin{align}
    \begin{split}
    u(t+\tau,x,a) = u(t,x,a) &+ \frac{\theta}{2}\left(u(t,x-\delta,a) + u(t,x+\delta,a)  - 2 u(t,x,a)  \right) \\
    & + \frac{\phi}{2}\left(\left[ c(x) - c(x-\delta) \right]_+ u(t,x-\delta,a) + \left[ c(x) - c(x+\delta) \right]_+ u(t,x+\delta,a) \right)  \\
    & - \frac{\phi}{2}\left( \left[c(x-\delta) - c(x)\right]_+ + \left[c(x+\delta) - c(x)\right]_+  \right) u(t,x,a) \\
    & + \psi_+ \left( 1 - \frac{a-\alpha}{A_{\text{max}}} \right) \mathds{1}_A(x) u(t,x,a-\alpha) - \psi_+ \left( 1 - \frac{a}{A_{\text{max}}} \right) \mathds{1}_A(x) u(t,x,a) \\
    & + \psi_- \left(\frac{a+\alpha}{A_{\text{max}}} \right) \left[1-\mathds{1}_A(x)\right] u(t,x,a+\alpha) - \psi_- \left(\frac{a}{A_{\text{max}}} \right) \left[1-\mathds{1}_A(x)\right] u(t,x,a).
    \end{split}
\end{align}
Subtracting $u(t,x,a)$ and dividing by $\tau$, we obtain:
\begin{align}
    \begin{split}
    \frac{u(t+\tau,x,a) - u(t,x,a)}{\tau} =  & \frac{\theta}{2\tau}\left(u(t,x-\delta,a) - 2 u(t,x,a) + u(t,x+\delta,a) \right)  \\
    & + \frac{\phi}{2\tau}\left(\left[ c(x) - c(x-\delta) \right]_+ u(t,x-\delta,a) + \left[ c(x) - c(x+\delta) \right]_+ u(t,x+\delta,a) \right)  \\
    & - \frac{\phi}{2\tau}\left( \left[c(x-\delta) - c(x)\right]_+ + \left[c(x+\delta) - c(x)\right]_+  \right) u(t,x,a) \\
    & + \frac{\psi_+}{\tau} \mathds{1}_A(x) \left( \left( 1 - \frac{a-\alpha}{A_{\text{max}}} \right)  u(t,x,a-\alpha) - \left( 1 - \frac{a}{A_{\text{max}}} \right)  u(t,x,a) \right) \\
    & + \frac{\psi_-}{\tau}  \left[1-\mathds{1}_A(x)\right] \left( \left(\frac{a+\alpha}{A_{\text{max}}} \right) u(t,x,a+\alpha) - \left(\frac{a}{A_{\text{max}}} \right) u(t,x,a) \right).
    \end{split}
\end{align}

By Taylor expanding about $x \pm \delta$, and taking the limit as $\tau,\delta \rightarrow 0$, then:
\begin{align*}
    \frac{\theta \delta^2 }{2\tau} \rightarrow D \in \mathbb{R}_+, \text{ as } \tau,\delta \rightarrow 0,
\end{align*}
the term $\frac{\theta}{2\tau}\left(u(t,x-\delta,a) - 2 u(t,x,a) + u(t,x+\delta,a) \right)$ can be written as:
\begin{align*}
    \frac{\theta}{2\tau}\left(u(t,x-\delta,a) - 2 u(t,x,a) + u(t,x+\delta,a) \right) & = \frac{\theta \delta^2 }{2\tau}\left( \frac{u(t,x-\delta,a) - 2 u(t,x,a) + u(t,x+\delta,a)}{\delta^2} \right), \\
    & = D \frac{\partial^2 u}{\partial x^2} + h.o.t. .
\end{align*}

Similarly, the terms governing chemotaxis can be written as:
\begin{align*}
    \frac{\phi}{2\tau} \Big( & \left[ c(x) - c(x-\delta) \right]_+ u(t,x-\delta,a) + \left[ c(x) - c(x+\delta) \right]_+ u(t,x+\delta,a) \\
    & -  \left( \left[c(x-\delta) - c(x)\right]_+ + \left[c(x+\delta) - c(x)\right]_+  \right) u(t,x,a) \Big), \\
= \frac{\phi}{2\tau} \Big( & \left[ c(x) - c(x-\delta) \right]_+ \left( u - \delta \frac{\partial u}{\partial x} + O(\delta^2) \right) + \left[ c(x) - c(x+\delta) \right]_+ \left( u + \delta \frac{\partial u}{\partial x} + O(\delta^2) \right) \\
    & -  \left[c(x-\delta) - c(x)\right]_+ u - \left[c(x+\delta) - c(x)\right]_+ u \Big), \\
= \frac{\phi}{2\tau} \Bigg( & 
\left\{\left[ c(x) - c(x-\delta) \right]_+ - \left[ -c(x) + c(x-\delta) \right]_+ \right\} u + \left\{\left[ c(x) - c(x+\delta) \right]_+ - \left[ -c(x) + c(x+\delta) \right]_+ \right\} u \\
& + \left\{ - \delta \left[ c(x) - c(x-\delta) \right]_+ \frac{\partial u}{\partial x} + \delta \left[ c(x) - c(x+\delta) \right]_+ \frac{\partial u}{\partial x}\right\} + O(\delta^2) \Bigg),\\
= \frac{\phi}{2\tau} \Bigg( & 
\left\{\left[ c(x) - c(x-\delta) \right]_+ - \left[ -c(x) + c(x-\delta) \right]_+ \right\} u + \left\{\left[ c(x) - c(x+\delta) \right]_+ - \left[ -c(x) + c(x+\delta) \right]_+ \right\} u \\
& + \delta^2 \frac{\partial u}{\partial x} \left\{ \left[ \frac{c(x) - c(x-\delta)}{\delta} \right]_+ - \left[ \frac{c(x) - c(x+\delta)}{\delta} \right]_+  \right\} + O(\delta^2) \Bigg),\\
= \frac{\phi}{2\tau} \Bigg( & 
\left\{\left[ c(x) - c(x-\delta) \right]_+ - \left[ -c(x) + c(x-\delta) \right]_+ \right\} u + \left\{\left[ c(x) - c(x+\delta) \right]_+ - \left[ -c(x) + c(x+\delta) \right]_+ \right\} u \\
& + \delta^2 \frac{\partial u}{\partial x} \left\{ \left[ \frac{c(x) - c(x-\delta)}{\delta} \right]_+ - \left[ \frac{c(x) - c(x+\delta)}{\delta} \right]_+ \right\} + O(\delta^2) \Bigg),\\
= \frac{\phi}{2\tau} \Bigg( & 
\left\{\left[ c(x) - c(x-\delta) \right]_+ - \left[ -c(x) + c(x-\delta) \right]_+ \right\} u + \left\{\left[ c(x) - c(x+\delta) \right]_+ - \left[ -c(x) + c(x+\delta) \right]_+ \right\} u \\
& + \delta^2 \frac{\partial u}{\partial x} \left\{ \left[ \frac{\partial c}{\partial x} \right]_+  - \left[- \frac{\partial c}{\partial x} \right]_+ \right\} + O(\delta^2) \Bigg).
\end{align*}
Using the property that $(c)_+ - (-c)_+ = c$:
\begin{align*}
=& \frac{\phi}{2\tau} \Bigg(
\left[ c(x) - c(x-\delta) \right] u + \left[ c(x) - c(x+\delta) \right] u - \delta^2 \frac{\partial u}{\partial x} \frac{\partial c}{\partial x} + O(\delta^2) \Bigg), \\
=& \frac{\phi}{2\tau} \Bigg(
\delta^2 u \left[ \frac{2 c(x) - c(x-\delta) - c(x+\delta)}{\delta^2} \right] - \delta^2 \frac{\partial u}{\partial x} \frac{\partial c}{\partial x} + O(\delta^2) \Bigg),\\
=& \frac{\phi \delta^2 }{2\tau} \Bigg( -u \frac{\partial^2 c}{\partial x^2} - \frac{\partial u}{\partial x} \frac{\partial c}{\partial x}\Bigg)  + O(\delta^2) ,\\
=& - \chi \frac{\partial}{\partial x}\Bigg(u \frac{\partial c}{\partial x} \Bigg)  + h.o.t.,
\end{align*}
since as $\tau,\delta \rightarrow 0$:
\begin{align*}
    \frac{\phi \delta^2 }{2\tau} \rightarrow \chi \in \mathbb{R}_+, \text{ as } \tau,\delta \rightarrow 0.
\end{align*}
For the final term, for simplicity, we rewrite as:
\begin{align*}
    & \frac{\psi_+}{\tau} \mathds{1}_A(x) \left[ \left( 1 - \frac{a-\alpha}{A_{\text{max}}} \right)  u(t,x,a-\alpha) - \left( 1 - \frac{a}{A_{\text{max}}} \right)  u(t,x,a) \right] \\
    & + \frac{\psi_-}{\tau}  \left[1-\mathds{1}_A(x)\right] \left[ \left(\frac{a+\alpha}{A_{\text{max}}} \right) u(t,x,a+\alpha) - \left(\frac{a}{A_{\text{max}}} \right) u(t,x,a) \right], \\
= & \frac{\psi_+}{\tau} \mathds{1}_A(x) \left[ - \alpha \frac{\partial}{\partial a} \left\{ \left( 1 - \frac{a}{A_{\text{max}}} \right)  u \right\}  + O(\alpha^2) \right] + \frac{\psi_-}{\tau}  \left[1-\mathds{1}_A(x)\right] \left[ \alpha \frac{\partial}{ \partial a} \left\{ \left(\frac{a}{A_{\text{max}}} \right) u \right\} + O(\alpha^2) \right], \\
= & - \frac{\psi_+ \alpha}{\tau} \mathds{1}_A(x)  \frac{\partial}{\partial a} \left\{ \left( 1 - \frac{a}{A_{\text{max}}} \right)  u \right\}   + \frac{\psi_- \alpha }{\tau}  \left[1-\mathds{1}_A(x)\right] \frac{\partial}{ \partial a} \left\{ \left(\frac{a}{A_{\text{max}}} \right) u \right\} + O(\alpha^2), \\
= & - \frac{\partial}{\partial a} \left\{ \left[ \frac{\psi_+ \alpha}{\tau}  \left( 1 - \frac{a}{A_{\text{max}}} \right)  \mathds{1}_A(x)   - \frac{\psi_- \alpha }{\tau} \left(\frac{a}{A_{\text{max}}} \right) \left[1-\mathds{1}_A(x)\right] \right] u \right\} + O(\alpha^2),\\
= & - \frac{\partial}{\partial a} \left\{ \left[ \mu_+  \left( 1 - \frac{a}{A_{\text{max}}} \right)  \mathds{1}_A(x)   - \mu_- \left(\frac{a}{A_{\text{max}}} \right) \left[1-\mathds{1}_A(x)\right] \right] u \right\} + h.o.t.,
\end{align*}
since as $\tau, \alpha \rightarrow 0$:
\begin{align*}
    \frac{\psi_+ \alpha }{\tau} \rightarrow \mu_+ \in \mathbb{R}, \quad \text{and} \quad \frac{\psi_- \alpha }{\tau}  \rightarrow \mu_- \in \mathbb{R}, \quad \text{as } \tau,\alpha \rightarrow 0.
\end{align*}
Since the term $\frac{u(t+\tau,x,a) - u(t,x,a)}{\tau}$ can simply be written as $\frac{\partial u }{\partial t}$ as $\tau \rightarrow 0$, we arrive at the PS-PDE:
\begin{align}
    \frac{\partial u }{\partial t} = D \frac{\partial^2 u}{\partial x^2}  - \chi \frac{\partial}{\partial x}\Bigg(u \frac{\partial c}{\partial x} \Bigg) - \frac{\partial}{\partial a} \left\{ \left[ \mu_+  \left( 1 - \frac{a}{A_{\text{max}}} \right)  \mathds{1}_A(x)   - \mu_- \left(\frac{a}{A_{\text{max}}} \right) \left[1-\mathds{1}_A(x)\right] \right] u \right\}.
\end{align}

\section{Multiple Dendritic cell activation sites in two-dimensions \label{si:multiple_DCs}}
In this section, we show how increasing the length scaling of the activation region, via higher dendritic cell clustering, is a predictor of how many T cells will become activated.
Here, we fix the parameter values $\mu_+ = 0.5$, $\mu_- = 0.5$, $A_{\text{max}} = 50$, and the movement parameters $D = 0.5$, and $\chi = 1.0$.

\begin{figure}[H]
    \begin{subfigure}[b]{0.99\textwidth}
        \centering
        \includegraphics[width=\linewidth]{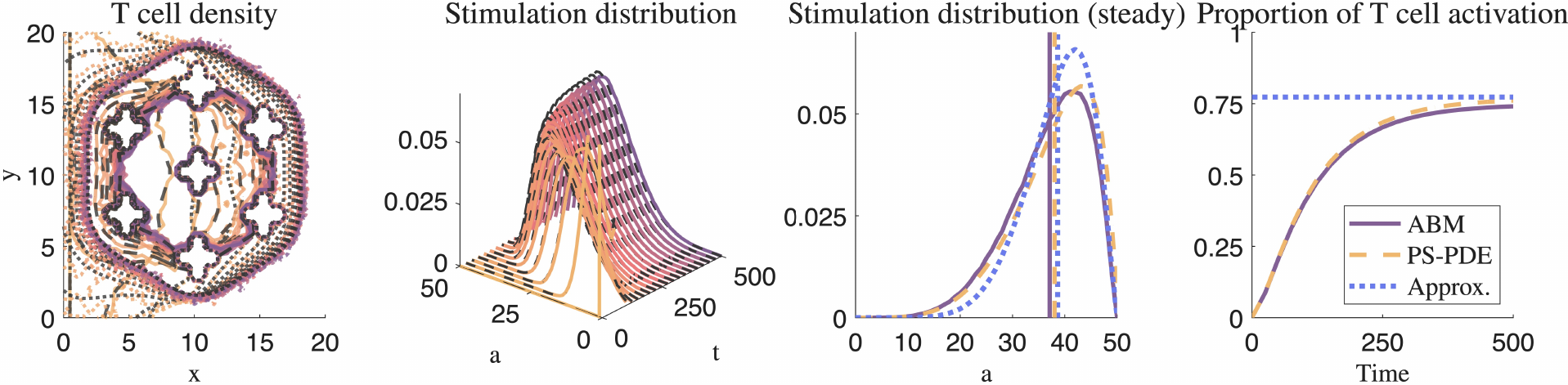}
        \caption{\label{fig:7Cells}}
    \end{subfigure}

    \begin{subfigure}[b]{0.99\textwidth}
        \centering
        \includegraphics[width=\linewidth]{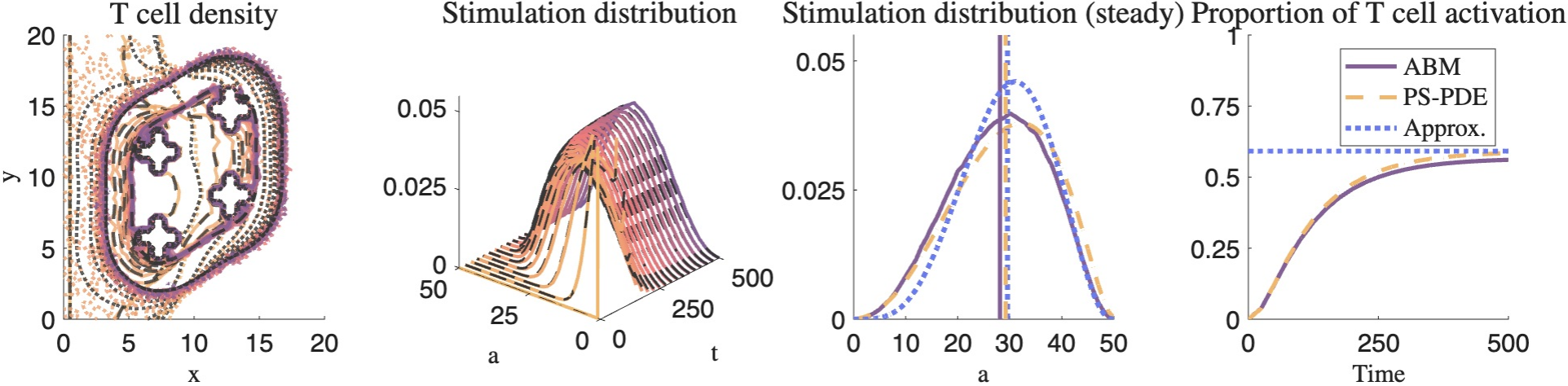}
        \caption{\label{fig:4Cells}}
    \end{subfigure}

    \begin{subfigure}[b]{0.99\textwidth}
        \centering
          \includegraphics[width=\linewidth]{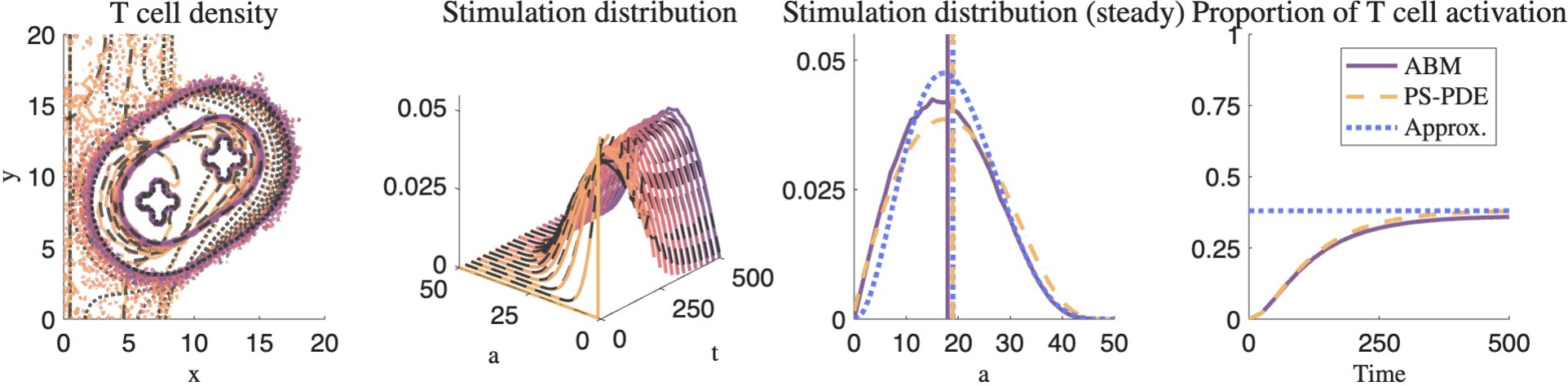}
        \caption{\label{fig:2Cells}}
    \end{subfigure}
    
    \caption{Examples of skewed high T cell activation (top row), mid level T cell activation (middle row) and skewed low T cell activation (bottom row). Left column show the contour plots of T cell density for PS-PDE model (black) and ABM (coloured from yellow/light to purple/dark to indicate time), with T cell density contours of $u_1 = 10^{-3}$ (dashed) and $u_2 = 5\times10^{-4}$ (dotted). Middle left column show the Stimulation distribution for the PS-PDE model (dashed black) and the ABM (solid coloured from yellow/light to purple/dark to indicate time), with time. Middle right column shows the steady-state stimulation distribution for the ABM (solid purple), PS-PDE model (dashed yellow) and the analytic approximation (dotted blue), and corresponding mean values (vertical corresponding lines). Right column shows the proportion of T cell activation over time for the ABM (solid purple), PS-PDE model (dashed yellow) and the analytic approximation (dotted blue). We observe close matches for the ABM and PS-PDE model in T cell density and stimulation distribution. However, for the steady-state stimulation distribution, we observe a mismatch in the distribution when comparing the analytic approximation to the ABM/PS-PDE model, but agreement in the mean level of stimulation. Model parameters: $\chi=1$, $D=0.5$, $\mu_- = 0.5$, $\mu_+ = 0.5$, and $A_{\text{max}} = 50$.}
    \label{fig:2D_cases}
\end{figure}

Figure \ref{fig:2D_cases} shows the results for three cases: skewed high (Figure \ref{fig:7Cells}), mid level (Figure \ref{fig:4Cells}) and skewed low  T cell activation (Figure \ref{fig:2Cells}). These different levels of T cell activation are achieved by increasing the area of activation, via increasing the number of DCs in the cluster, with the results directly comparable to the 1D case presented in Figure \ref{fig:1D_cases_123}. We observe that instances with more larger Dendritic cell clusters (i.e. more DCs) result in a higher level of T cell activation.
However, from these results, we are not able to derive insight into how a particular topology influences the time-scale, or how the degree of clustering influences T cell activation. This is because, in these instances, the number of DCs (equivalently interpreted as the density of DCs) within the domain is also changing.

\section{Dimensional scaling and numerical scheme}\label{si:numerics}
To simulate the system, we scale dimensions as:
\begin{itemize}
    \item \textbf{space:} as $1$ spatial unit $=$ $4 \mu$ meters,
    \item \textbf{time:} as $1$ temporal unit $=$ $1$ minute.
\end{itemize}

\subsection{ABM numerical scheme}
To solve the ABM, following initialisation of the system, we perform the following algorithm:
\begin{algorithm}
\caption{Numerical scheme to update ABM}
\begin{algorithmic}[1]
\STATE Initialise $x_0$
\FOR{time $= 1$ to $T_{\text{final}}$}
    \STATE Compute random movement probabilities $\mathbf{R}$, per Equations (\ref{eq:rand_prob})
    \STATE Update spatial positions $\Vec{x}$
    \STATE Compute chemotaxis movement probabilities $\mathbf{T}$, per Equation (\ref{eq:taxis_prob})
    \STATE Update spatial positions $\Vec{x}$
    \STATE Check for spatial boundary conditions
    \STATE Compute stimulation probabilities $\mathbf{P}$, per Equations (\ref{eq:stim_up}) and (\ref{eq:stim_down}). 
    \STATE Update stimulation $\Vec{a}$
    \STATE Check for stimulation boundary conditions
\ENDFOR
\STATE \textbf{return} $(\Vec{x},\Vec{a})$
\end{algorithmic}
\end{algorithm}

\subsection{PS-PDE numerical scheme}
To solve the PS-PDE, following initialisation of the model and definition of the boundary and initial conditions, we implement a finite volume scheme for the spatial dynamics, and an upwinding scheme for the stimulation dynamics. In one spatial dimension, if we let $u^k_{i,j} = u(t_k,x_i,a_j)$, then:
\begin{align*}
    u^{k+1}_{i,j} &= u^{k}_{i,j} + dt( J_E - J_W + J_A),\\
    &= u^{k}_{i,j} + \Delta t \Bigg\{ \left[ D \frac{\left( u^{k}_{i+1,j} - u^{k}_{i,j}\right)}{\Delta x} - \chi\frac{\left( u^{k}_{i+1,j} + u^{k}_{i,j}\right)}{\Delta x}\frac{\left( C_{i+1} - C_{i}\right)}{2}\right]/\Delta x \\
  & \quad \qquad \qquad - \left[ D \frac{\left( u^{k}_{i,j} - u^{k}_{i-1,j}\right)}{\Delta x} - \chi\frac{\left( u^{k}_{i,j} + u^{k}_{i-1,j}\right)}{\Delta x}\frac{\left( C_{i} - C_{i-1}\right)}{2}\right]/ \Delta x\\
  & \quad \qquad \qquad + \left[ f_{a,j}^{-} u^{k}_{i,\Delta-} - f_{a,j}^{+} u^{k}_{i,\Delta+} \right]/ \Delta a \Bigg\},
\end{align*}

where:
\begin{align*}
    f_a^+ &= \mu_+\left(1-\frac{a_{+1/2}}{A_{\text{max}}} \right) \mathds{1}_{\text{A}_{i}} - \mu_- \frac{a_{+1/2}}{A_{\text{max}}} \left(1- \mathds{1}_{\text{A}_{i}}\right),\\
    f_a^- &= \mu_+\left(1-\frac{a_{-1/2}}{A_{\text{max}}} \right) \mathds{1}_{\text{A}_{i}} - \mu_- \frac{a_{-1/2}}{A_{\text{max}}} \left(1- \mathds{1}_{\text{A}_{i}}\right),\\
    u^{k}_{i,\Delta+} &= \begin{cases}
        u^{k}_{i,j}, \quad \text{for } f_a^+ > 0,\\
        u^{k}_{i,j+1}, \quad \text{for } f_a^+ \leq 0,
    \end{cases}\\
    u^{k}_{i,\Delta-} &= \begin{cases}
        u^{k}_{i,j}, \quad \text{for } f_a^- \leq 0,\\
        u^{k}_{i,j-1}, \quad \text{for } f_a^- > 0,
    \end{cases}\\
    a_{-1/2} &= \left(a_{j+1} + a_j \right)/2, \\
    a_{+1/2} &= \left(a_{j} + a_{j-1} \right)/2.
\end{align*}

\end{document}